\documentclass{jfm}
\usepackage{graphicx}
\usepackage{newtxtext}
\usepackage{newtxmath}
\usepackage{natbib}
\usepackage{hyperref}
\hypersetup{
    colorlinks = true,
    urlcolor   = blue,
    citecolor  = blue,
}

\usepackage{xcolor}
\usepackage{etoolbox}
\usepackage{tabularx}
\usepackage{tikz}
\usetikzlibrary{positioning}
\usepackage{subcaption}
\usepackage{comment}
\definecolor{tabblue}{RGB}{8,73,145}
\definecolor{tabgreen}{RGB}{0,100,40}
\definecolor{tabpurple}{RGB}{79,30,139}
\definecolor{tabred}{RGB}{151,11,19}
\definecolor{taborange}{RGB}{254,140,59}
\definecolor{tabbrown}{RGB}{140,86,75}

\title{Mean flow scaling in stably stratified temporally developing turbulent boundary layers}
\author{Baptiste Hardy\aff{1} \corresp{\email{baptiste.hardy@uclouvain.be}}, Pedro Costa \aff{2}}
\affiliation{\aff{1} Université catholique de Louvain, Institute of Mechanics, Materials and Civil Engineering, Place du Levant, 2, 1348 Louvain-la-Neuve, Belgium
\aff{2}Process \& Energy Department, Faculty of Mechanical Engineering, Delft University of Technology, Leeghwaterstraat 39, 2628~CB, Delft, The Netherlands}

\begin{document}

\maketitle

\begin{abstract}

Stably stratified wall-bounded turbulence governs the dynamics of many environmental and engineering flows. A key challenge is characterizing how stratification modifies mean and turbulent profiles. Monin--Obukhov similarity theory (MOST) is the dominant modelling framework, although it has rarely been rigorously validated against well-controlled direct numerical simulation (DNS) data over a wide range of stratification levels. In this study, we exploit the temporally developing turbulent boundary layer (TTBL) framework to investigate stratified turbulent boundary layers from the weakly stable to the very stable regime, spanning a range of Reynolds and Richardson numbers, and isolating the effects due to buoyancy from other mechanisms such as flow rotation. We demonstrate that the TTBL set-up faithfully reproduces classical similarity theory results and that surface-based scaling of the mean velocity gradient holds over a wider range of $z/L$ ($L$ being the Obukhov length) than previously reported. This result is attributed to the similar decay rate of turbulent shear stress and heat flux in this canonical flow. Next, we show that, as stratification intensifies, the intercept of the mean velocity profile increases, until the separation of scales required for a logarithmic region to exist can no longer be sustained. We propose an empirical closure for this intercept shift in terms of the Reynolds number based on the Obukhov length. Finally, a simple damping of the MOST contribution to the mean velocity profile is proposed and validated, enabling accurate prediction of the wall friction coefficient ($C_f$) across the investigated regimes. %

\end{abstract}

\section{Introduction}

In various environmental phenomena and engineering applications, a turbulent boundary layer (TBL) is subject to a negative surface heat flux. The thermal stratification that results from the positive temperature gradient stabilizes the TBL and strongly attenuates turbulent motions. The modelling of the stably stratified atmospheric boundary layer (ABL), in particular, has been a very active area of research over the past decades. A central framework in this context is the Monin--Obukhov similarity theory (MOST) \citep{Monin-Obukhov1954}, which relates dimensionless mean gradients and turbulence statistics in the near-wall region to the competition between mechanical shear and buoyancy through the Obukhov length scale. MOST and its later extensions \citep{nieuwstadt1984, holtslag1986,sorbjan2010} underpin many descriptions of stable wall-bounded flows, from geophysical boundary layers to engineering models. Field measurements \citep{businger1971, dyer1974, nieuwstadt1984, hunt1985} and laboratory experiments \citep{arya1969,ohya1997} provided the first systematic insights into the modification of the mean velocity profile and turbulence statistics when the ABL is subject to ground cooling, and supplied much of the evidence used to assess and calibrate these similarity theories. Recent laboratory-scale experimental studies by \cite{williams2017} and \cite{ding2024} revisited these classical ideas, exploring a wider range of stratification levels and assessing the effect of wall roughness on scaling laws and turbulence stability criteria.

The complexity of the stably stratified ABL arises from the variety of regimes it can exhibit, depending on the intensity of the ground cooling. Similarity theories were typically developed in the weakly stable regime, where turbulence is still shear-dominated, continuous, and quasi-steady, while being progressively suppressed by stratification through buoyancy effects. However, under stronger stratification conditions -- the very stable regime -- turbulence becomes intermittent, characterized by periods of weak fluctuations, and may ultimately collapse \citep{mahrt1999, mahrt2014}. This regime dependence makes the stable atmospheric boundary layer difficult to describe, predict, and model within a single unified framework, and remains the subject of active research \citep{heisel2023, bon2024, shen2024}.

Numerical simulations have also been widely employed to study stably stratified TBLs. \cite{coleman1992} reported the first direct numerical simulation (DNS) of a stably stratified turbulent Ekman layer where the flow is driven by a geostrophic wind forcing. \cite{shah2014} as well as \cite{ansorge2014} used the computational set-up proposed by \citet{coleman1992} to investigate turbulence statistics for a wider range of Reynolds numbers and stratification levels, from the neutral to the very stable regime. %
\cite{nieuwstadt_2005}, \cite{flores2011} and \cite{atoufi2019} performed direct numerical simulations of stratified open-channel flows while ignoring Earth's rotation, therefore being more representative of a nocturnal surface layer than of the full atmospheric boundary layer. These studies imposed a constant negative heat flux on the bottom wall to represent radiative cooling, while \citet{coleman1992}, \citet{shah2014} and \citet{ansorge2014} prescribed a fixed temperature. Despite these advancements, DNS remains unreachable for realistic atmospheric conditions, and large-eddy simulation (LES) must be employed. In this context, modelling subgrid-scale stability effects and land-surface fluxes remains challenging because eddy sizes decrease substantially even in the weakly stable regime, while turbulence becomes intermittent in the very stable regime \citep{mahrt1998}. 

Most numerical studies mentioned above relied on the homogeneity and quasi-steady state of the stably stratified TBL, and therefore computed Reynolds-averaged quantities using both temporal and spatial averages along the homogeneous directions. Conversely, space-developing simulations of stratified TBLs are quite rare: \cite{hattori2007} performed DNS of stable and unstable spatially developing TBLs in relatively small domains, while \cite{tomas2016} performed spatially developing LES of a stable TBL over roughness elements mimicking an urban environment. These studies employed periodic precursor simulations to generate turbulent inflow conditions for the spatially developing simulation, which is complex and computationally expensive. Synthetic inflow approaches, such as the synthetic-eddy method of \cite{jarrin2006}, could in principle provide an alternative, but they are even more challenging to adopt in the present context: in addition to realistic velocity fluctuations at inflow, stable stratification effects also require a consistent prescription of buoyancy-affected anisotropy and turbulent heat-flux statistics.

To avoid these limitations, time-developing frameworks provide an interesting alternative. The temporally developing turbulent boundary layer (TTBL) has been thoroughly investigated by \cite{Kozul2016} in the neutral regime (i.e. in the absence of buoyancy effects). In the TTBL, the flow is driven by a wall moving at a prescribed velocity, analogous to the free-stream velocity in the spatially developing case, while the far field is at rest. The boundary layer grows in time, and turbulent statistics collected at a specific time of the simulation would be representative of a specific streamwise position in a spatially developing set-up. 
The streamwise and spanwise homogeneity of the flow makes the TTBL an attractive configuration for investigating fundamental TBL properties at reduced computational cost, as proven by recent studies \citep{kozul2020,cimarelli2024-1}. 
\citet{Kozul2016} showed that this computational set-up reproduces classical statistics of the spatially developing TBL (profiles of mean and turbulent quantities, skin friction), and further demonstrated that this set-up corresponds to the asymptotic limit of the spatially developing counterpart as the flow becomes locally parallel (i.e. at high-enough Reynolds number). As such, it provides a useful framework for studying developed wall-bounded turbulence. These observations have been further consolidated in subsequent studies \citep{kozul2020,cimarelli2024-1}. 

However, the validity of surface and local similarity scaling in a non-rotating, temporally developing stably stratified boundary layer has not been rigorously established. %
In this work, we perform direct numerical simulations of the stably stratified TTBL to address this gap, and investigate stability effects across a broad range of governing dimensionless parameters at a significantly lower cost than in spatially developing boundary layers. Our results show that the TTBL set-up recovers the classical Monin--Obukhov similarity results, and that, in the absence of flow rotation, surface scaling of the mean velocity gradient remains valid further from the surface than previously reported. We further derive an empirical closure relating the upward shift in the log-linear intercept to the Reynolds number based on the Obukhov length. Finally, we propose a model for the log-linear and outer parts of the mean velocity profile, as a function of the stratification level, enabling accurate prediction of the wall friction coefficient.

This manuscript is organized as follows. In Section \ref{section:theory}, we first recall the foundation of similarity theory for stratified turbulent boundary layers and we describe the numerical methodology. In Section \ref{section:setup}, we detail our computational set-up and DNS parameters. In Section \ref{section:results}, we present and discuss the results of our stably stratified TTBL simulations. Finally, in Section \ref{section:conclusion}, we summarize the main conclusions of this work.

\section{Theory and numerical methodology} 
\label{section:theory}

\subsection{Similarity theory}
\cite{Monin-Obukhov1954} laid the foundations of a similarity theory targeting the universal scaling of the flow in the surface layer of the stratified ABL. Based on dimensional analysis, these authors argued that the two relevant length scales in the surface layer are the distance to the wall ($z$) and the Obukhov length \citep{obukhov1946}, defined as 
\begin{equation}
L= - \frac{u_\tau^3}{\kappa \beta g q_w}, 
\label{eq:ObukhovL}
\end{equation}
where $u_\tau $ is the wall friction velocity, $\kappa = 0.41$ is the von-Kármán constant, $\beta = 1/T_0$ is the thermal expansion coefficient (with $T_0$ being the reference base flow temperature in the Boussinesq approximation), $g$ is the gravitational acceleration, and %
$q_w$ is the kinematic surface heat flux. %
The Obukhov length characterizes the distance from the wall at which buoyancy effects become dominant over shear-induced turbulence production. Assuming stationary and horizontally homogeneous turbulence, MOST states that dimensionless flow statistics in the constant-flux surface layer of a stratified TBL depend only on the ratio $\zeta = z/L $. For the mean velocity and temperature, this is usually expressed through similarity functions $\phi_m$ and $\phi_h$ as
\begin{align}
    \frac{\kappa z}{u_\tau}\frac{\partial \bar u}{\partial z} = \phi_m(\zeta), \quad \textrm{and} \quad  
    \frac{\kappa z}{\Theta_\tau}\frac{\partial \bar \Theta}{\partial z} = \phi_h \left(\zeta\right) \label{eq:MOST}, 
\end{align}
where $u$ is the streamwise velocity component, $\Theta = T - T_0$ is the temperature difference with respect to the Boussinesq base state, and $\Theta_\tau = -q_w/u_\tau$ is the wall friction temperature. In this work, the bar symbol indicates Reynolds-averaging, while primed quantities denote fluctuations with respect to this average. In weakly stable conditions, the Businger--Dyer linear relations proposed for $\phi_m$ and $\phi_h$ \citep{webb1970, businger1971, dyer1974} have been shown to fit field data reasonably well:
\begin{align}
    \phi_m(\zeta) &= 1 + \beta_m \zeta, \label{eq:phi_m}\\
    \phi_h(\zeta) &= \alpha_h + \beta_h \zeta, \label{eq:phi_h}
\end{align}
where the values of parameters $\beta_m$, $\beta_h$ and $\alpha_h$ are discussed later. 
In strongly stable conditions (i.e.\ large $\zeta$), the largest turbulence length scale -- and, consequently, the mixing length -- is capped by stratification. As a result, the mean velocity gradient becomes independent of the height, leading to the well-established $z$-less scaling region of the stably stratified boundary layer \citep{wyngaard1973}. Note that this phenomenon is captured by Eqs.~\eqref{eq:phi_m} and \eqref{eq:phi_h}, as it can be easily observed by plugging the definitions of $\phi_{m/h}$ into Eq.~\eqref{eq:MOST} and taking the limit for $\zeta \gg 1$. Later studies \citep{beljaars1991,cheng2005} reported a slower (i.e. non-linear) evolution of dimensionless groups with $\zeta$ for larger stratification levels and, in some cases, a flattening of the similarity functions. Using Beljaars--Holtslag nonlinear similarity relations, \cite{sharan2003} showed that the correct $z/L$ ratio -- and hence surface fluxes -- could be recovered from the bulk Richardson number measured in weak-wind stable conditions, whereas the linear Businger--Dyer relations systematically underestimate the Obukhov length once the bulk Richardson number exceeds a critical threshold of about 0.2\footnote{Details about Richardson number definitions and their connection to similarity theory are provided in Appendix \ref{appendix:richardson}.}.

An important milestone in ABL modelling was achieved by \cite{nieuwstadt1984}, who established that mean gradients (and other flow statistics) in the stratified ABL should be scaled by local turbulent fluxes rather than surface values, thereby extending the validity of the similarity theory beyond the near-wall region where turbulent and surface fluxes are approximately equal, namely the constant-flux surface layer. This framework is now referred to as local scaling theory, in contrast to the surface-layer-based (or global) scaling of classical MOST discussed above. The local Obukhov length is then defined as
\begin{equation}
    \Lambda(z)= -\frac{\tau(z)^{3/2}}{\kappa g \beta q(z)},
\end{equation}
where $\tau = - \overline{u'w'} = u_*^2$ is the turbulent shear stress, $q = \overline{w'\Theta'} = -u_*\Theta_*$ is the turbulent heat flux (negative under stable stratification) and $w$ is the wall-normal velocity component. The gradient similarity functions in local scaling are hence written as $\phi_{m,l} = (\kappa z)/ {u_*} \, \partial\bar u/\partial z$ and $\phi_{h,l} = (\kappa z)/ {\Theta_*}\,\partial\bar \Theta/\partial z $, and can be expressed as functions of the locally scaled wall distance $\zeta_l = z/\Lambda$, using closures of the same functional form as Eqs. \eqref{eq:phi_m} and $\eqref{eq:phi_h}$.

\subsection{Governing equations and boundary conditions} 
The computational set-up of this study is based on the temporally developing turbulent boundary layer framework described by \citet{Kozul2016}. The incompressible Navier--Stokes equations are solved with the Oberbeck--Boussinesq approximation to account for the effects of buoyancy on the flow. The governing equations are as follows: 
\begin{align}
    \nabla \cdot \mathbf U & = 0, \label{eq:NS1}\\
    \frac{\partial \mathbf U}{\partial t} + (\mathbf U \cdot \nabla) \mathbf U & = -\nabla P + \nu \nabla ^2 \mathbf U - \beta \Theta \mathbf g, \label{eq:NS2}\\
     \frac{\partial  \Theta }{\partial t} + (\mathbf U \cdot \nabla)  \Theta & = \alpha \nabla ^2 \Theta, \label{eq:NS3}
\end{align}
where $\mathbf U $ is the flow velocity in the domain, $P$ is the kinematic pressure, $\nu$ is the kinematic viscosity, $\alpha$ is the thermal diffusivity and  %
$\mathbf{g} = -g \mathbf{e}_z$ is the gravitational acceleration. %
Eqs.~\eqref{eq:NS1} to \eqref{eq:NS3} are integrated in time over a domain $(L_x,L_y,L_z)$ with a fractional-step method and Wray's third-order low-storage Runge--Kutta scheme \citep{Wray1990minimal}. The momentum and temperature diffusion terms are treated implicitly along the wall-normal direction using a Crank--Nicolson scheme to avoid the diffusion stability constraint when the grid is refined near the wall. 

In a TTBL framework, the flow is driven by a prescribed streamwise velocity $U_w$ at the wall, while the far field is at rest. %
At the upper boundary, a free-slip/no-penetration boundary condition is prescribed for the velocity field, and an adiabatic condition is imposed on temperature. The vertical extent of the domain has to be sufficiently large in order for the boundary-layer thickness $\delta$ to remain sufficiently small with respect to the height of the domain over the duration of the simulation, thereby preventing the upper boundary from altering the flow. Typically, the final thickness of the TBL, $\delta_f$, should not exceed one third of the height of the domain \citep{Kozul2016}.
Periodic boundary conditions are used along the homogeneous streamwise ($x$) and spanwise ($y$) directions. The boundary conditions in the wall-normal direction are given by
\begin{align}
     U = U_w, \quad V = 0, \quad W  = 0, \quad \frac{\partial \Phi}{\partial z} = 0, \quad \Theta = \Theta_w \quad  & \textrm{ at } z=0,  \label{eq:BC1}\\
     \frac{\partial U}{\partial z} = 0, \quad \frac{\partial V}{\partial z} = 0, \quad W  = 0, \quad \frac{\partial \Phi}{\partial z} = 0, \quad \frac{\partial \Theta}{\partial z} = 0 \quad  & \textrm{ at } z=L_z,  \label{eq:BC2}
\end{align}
where $\Phi$ is the pressure correction field, the solution of the Poisson equation in the fractional-step algorithm. The system of Eqs. \eqref{eq:NS1} to \eqref{eq:NS3} is solved with the boundary conditions \eqref{eq:BC1} and \eqref{eq:BC2} using the FFT-based and GPU-accelerated finite-difference solver CaNS \citep{costa2018}.

In order to connect the DNS results obtained in the stratified TTBL with existing similarity theory and previous literature results, we report in the rest of this study the profiles of the velocity components $(u,v,w)$ as they would be obtained in a spatially developing TBL with free-stream velocity $U_w$ and zero velocity at the wall. Therefore, the velocity components $(u,v,w)$ are defined from the DNS flow field $\mathbf U = (U,V,W)$ as 
\begin{equation}
    u = U_w - U, \quad v = V, \quad w = W.
     \label{eq:new_var}
\end{equation}

\section{Simulation set-up} 
\label{section:setup}
The initialization of the flow is of major importance in temporally developing boundary layers since it defines the initial amount of momentum injected in the domain, and is therefore connected to the momentum thickness Reynolds number $Re_\theta = U_w \theta/\nu$ of the corresponding spatially developing TBL, where $ \theta = \int_0^\infty \frac{\bar U}{U_w}\left(1 - \frac{\bar U}{U_w}\right)dz.$
The strategy proposed by \cite{Kozul2016} is reproduced here: the flow velocity is initialized over a characteristic trip distance $d_0$ from the wall to mimic experimental studies in which turbulence is triggered by a wall-mounted trip wire. The initial velocity field is defined as 
$\mathbf U_0 = \bar U_0(z) \mathbf e_x + \mathbf U_0'(x,y,z)$, 
where the mean streamwise component of the flow $\bar U_0(z)$ is given by
\begin{equation}
\bar U_0(z) = \frac{U_w}{2}\left(1 + \tanh\left[\frac{d_0}{2\theta_{sl}}\left( 1 - \frac{z}{d_0}\right) \right] \right).
\end{equation}
The momentum thickness of the shear layer is set to $\theta_{sl} = 54 \nu/U_w$. This value ensures that the Kelvin--Helmholtz rollers that form at initial times are sufficiently small and will be quickly forgotten as time progresses.  To trigger instability in the flow, white noise $|U_{0,i}'| < 0.05 \, U_w $ (where $i = 1,2,3$) is added to the three velocity components. The temperature field is initialized to the base state temperature $T_0$, i.e. $\Theta = 0$. The nominal Reynolds and Richardson numbers are defined as 
\begin{equation}
    Re_0 = \frac{U_w d_0}{\nu}, \textrm{  and  } Ri_0 = -\frac{\beta g \Theta_w d_0}{U_w^2}, 
\end{equation}
respectively. %
Since $\Theta_w < 0$ with surface cooling, $Ri_0$ is defined positive for stable stratification. The Prandtl number $Pr = \nu/\alpha$ is set to $0.71$ for all cases. 

To cover a range of stratification levels, we performed DNS for three different nominal Richardson numbers and two nominal Reynolds numbers, as indicated in table \ref{tab:parameters}. 
We also report the maximum friction Reynolds number $Re_\tau = u_\tau \delta/\nu$ for each simulation, as well as the final values (denoted by a subscript `$f$') achieved by the bulk Richardson number $Ri_\delta = -\beta g \Theta_w \delta/U_w^2$ and friction Richardson number $Ri_\tau = -\beta g \Theta_w \delta/u_\tau^2$. Unlike in neutral conditions, the final friction Reynolds number in the stably stratified TBL may be lower than its maximum value during the simulation because of turbulence destruction by stratification. To connect the time evolution of bulk quantities in the TTBL with their streamwise evolution in the corresponding spatially developing TBL, \citet{Kozul2016} introduced the Reynolds number based on the equivalent streamwise displacement $Re_X = U_w X/\nu$, where $X = U_w t$.
The boundary-layer thickness $\delta$ is estimated as the distance to the wall where $\bar U$ falls below $0.01 U_w$, equivalently where $\bar u$ reaches $0.99 U_w$. %

The computational campaign was chosen to ensure sufficiently high $Re_\tau$ to satisfy scale separation and enable us to study the mean flow scaling laws with minimal modelling assumptions. %
The size of the domain is scaled by $\nu/U_w$ along each direction: $L_x U_w /\nu  \times  L_y U_w /\nu \times L_z U_w /\nu  =  600\,000 \times 200\,000 \times 80\,000$. Since stratification reduces boundary-layer growth, and the current domain height is suitable for neutral cases, all cases feature (conservatively) sufficient heights. Similarly, since stratification tends to dampen turbulence, a grid sufficiently fine to resolve the neutral case will also resolve the stably stratified one. Here, the grid consists of $(n_x, n_y, n_z) = (3072, 2048, 512)$ points in the streamwise, spanwise and wall-normal directions, respectively. The grid is stretched in the wall-normal direction to satisfy the condition $\Delta z_b^+ \lesssim 1$ at the bottom wall,  $\Delta x^+ \lesssim 10$ and $\Delta y^+  \lesssim 5$ along the two homogeneous directions; see table \ref{tab:parameters}, with the superscript '$^+$' denoting wall-unit scaling, i.e. normalization by $\nu/u_\tau$.

Given the problem's homogeneity along the streamwise and spanwise directions, Reynolds averages are computed by spatial averaging along them. However, to obtain fully converged statistics, an ensemble average over multiple realizations of the flow -- each initialized with different random noise -- would also be required. Since the results presented in this study correspond to a single realization, the lack of convergence can be slightly mitigated by applying a sliding time average to the spatially averaged quantities \citep{Kozul2016}, provided the time-averaging window $T_{av}$  remains small compared to the characteristic time scale over which statistics of the TTBL evolve. $T_{av}$ has here been fixed to 500 time steps for all cases and for simplicity. This value corresponds to a ratio $\chi_{av}= T_{av}/(\delta/u_\tau)$ of 0.055 and 0.043 at the end of cases A3 and B1, respectively. In comparison, \citet{Kozul2016} used a ratio $\chi_{av} = 0.5$, i.e. a time-averaging window about 10 times wider. We are therefore confident that the results we report here are not altered by the time-averaging procedure. 
We verified the validity of our approach by comparing the time-averaged and instantaneous statistics directly, and found that the latter exhibit additional scatter but no systematic bias with respect to the former.

\begin{table}
    \centering
    \begin{tabularx}{\textwidth}{XXXXXXXXXX}
        \hline
         Case & $Re_0$& $Ri_0$ & $Re_{\tau,\max}$ & $Ri_{\tau,f}$ & $Ri_{\delta,f}$  & $\Delta x^+_{\max}$ & $\Delta y^+_{\max}$ & $\Delta z^+_{b,\max}$ & \\[0.2cm]
         \hline
          N&  1000 & 0 & 2305 & - & - & 10.47 & 5.23 & 1.24 & \textcolor{black}{\rule{16pt}{4pt}} \\[0.2cm]
         A1&  1000 & 0.004 & 588& 328 & 0.115 & 10.36 & 5.18 & 1.23 & \textcolor{tabblue}{\rule{16pt}{4pt}} \\[0.2cm]
         A2&  1000 & 0.005 & 477& 525& 0.123 & 10.07 & 5.04 & 1.19& \textcolor{tabpurple}{\rule{16pt}{4pt}}\\[0.2cm]
         A3&  1000 & 0.006 & 417& 551 & 0.130& 10.06 & 5.03 & 1.19 & \textcolor{tabgreen}{\rule{16pt}{4pt}}\\[0.2cm]
         B1&  1500 & 0.004 & 827& 201& 0.099& 9.14 & 4.56 & 1.08 &  \textcolor{tabred}{\rule{16pt}{4pt}}\\[0.2cm]
         B2&  1500 & 0.005 & 691& 258& 0.110& 9.12 & 4.55 & 1.08 & \textcolor{taborange}{\rule{16pt}{4pt}}\\[0.2cm]
         B3&  1500 & 0.006 & 605& 348& 0.117& 9.13 & 4.56 & 1.08 & \textcolor{tabbrown}{\rule{16pt}{4pt}}\\[0.2cm]
        \hline
    \end{tabularx}
    \caption{Physical and numerical parameters for the present DNS campaign. Values of $\Delta x^+_{\rm max}$, $\Delta y^+_{\rm max}$ and $\Delta z^+_{\rm b, max}$ correspond to the peak values achieved at early times.}
    \label{tab:parameters}
\end{table}
\section{Results}
\label{section:results}
\subsection{Time evolution of bulk quantities}

\begin{figure}
\begin{minipage}[b]{0.48\linewidth}
\includegraphics[width=.95\textwidth]{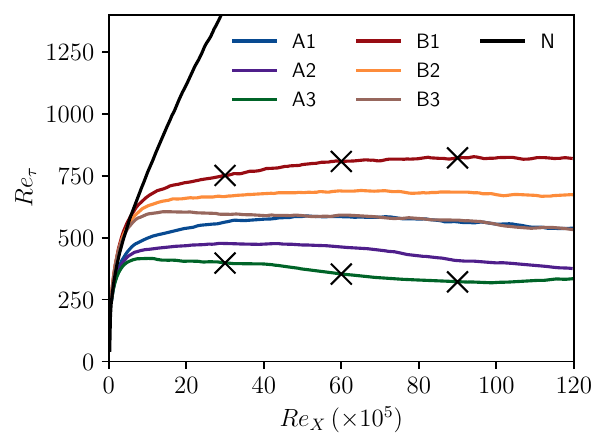}%
\begin{picture}(0,0)
    \put(-175,130){(a)}
\end{picture}
\end{minipage}
\hfill
\begin{minipage}[b]{0.48\linewidth}
\includegraphics[width=.95\textwidth]{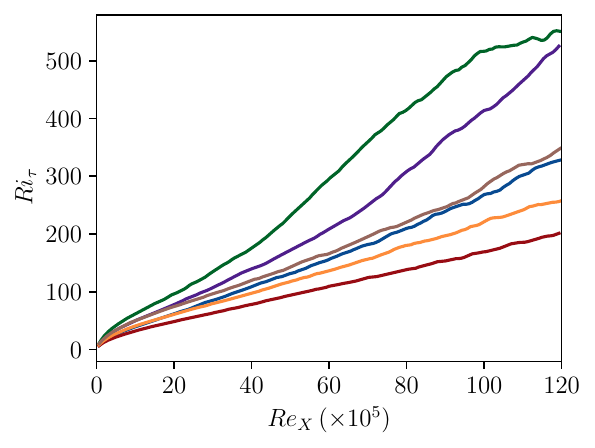}%
\begin{picture}(0,0)
    \put(-175,130){(b)}
\end{picture}
\end{minipage}
\begin{minipage}[b]{0.48\linewidth}
\includegraphics[width=.95\textwidth]{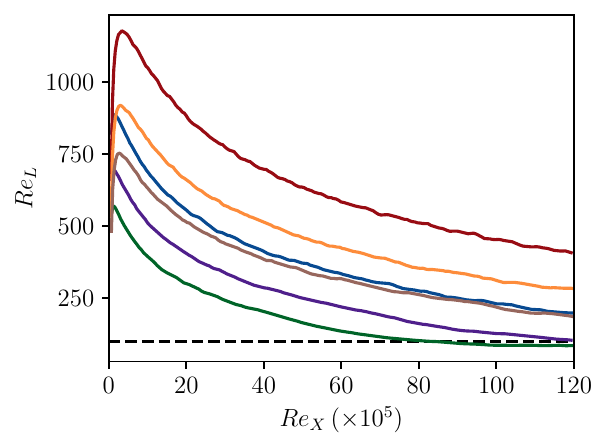}%
\begin{picture}(0,0)
    \put(-175,130){(c)}
\end{picture}
\end{minipage}
\hfill
\begin{minipage}[b]{0.48\linewidth}
\includegraphics[width=.95\textwidth]{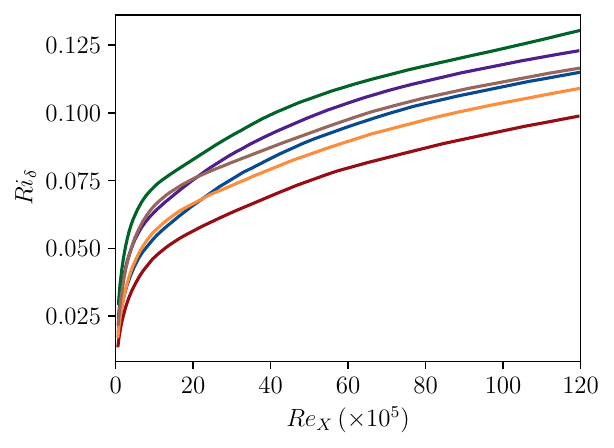}%
\begin{picture}(0,0)
    \put(-175,130){(d)}
\end{picture}
\end{minipage}
\caption{Temporal evolution of (a) the friction Reynolds number, (b) the friction Richardson number, (c) the Reynolds number based on Obukhov length  and (d) the bulk Richardson number for the different DNS cases. Crosses in (a) indicate the times of the snapshots shown in figure \ref{fig:tempfield_snapshots_yplane}. Dashed line in (c) corresponds to the relaminarization threshold $Re_L = 100$ proposed by \cite{flores2011}.}
\label{fig:time_evolving_quantities}
\end{figure}
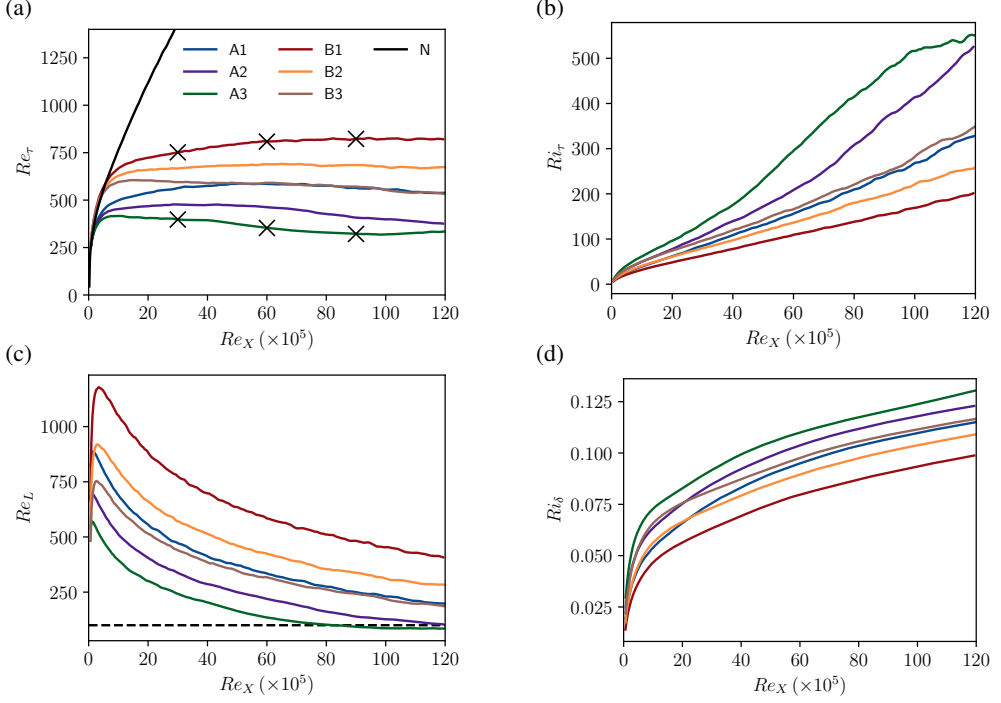

We first investigate the temporal evolution of dimensionless bulk quantities for the different cases listed in table \ref{tab:parameters}. Figure \ref{fig:time_evolving_quantities} shows the evolution of the friction Reynolds number $Re_\tau$, the friction Richardson number $Ri_\tau$, the Reynolds number based on the Obukhov length $Re_L = u_\tau L/\nu$ and the bulk Richardson number $Ri_\delta$ as a function of $Re_X$. As expected, the friction Reynolds number (figure \ref{fig:time_evolving_quantities}(a)) grows monotonically in the neutral case (the vertical scale was cropped for better readability). Conversely, this growth is capped in the stably stratified cases (A1--B3), with $Re_\tau$ eventually decaying for sufficiently long simulation times. In the stably stratified cases, the maximum value achieved by $Re_\tau$ increases with $Re_0$ and decreases with $Ri_0$, as stratification effects grow important sooner at higher $Ri_0$ and counteract shear forces more rapidly. On the contrary, the friction Richardson number (figure \ref{fig:time_evolving_quantities}(b)) keeps increasing steadily with time as a result of the continuous surface cooling. A flattening of the curve is observed for the most stratified case (A3) around $Re_X = 90 \times 10^5$. At this stage, stratification effects are strong enough to generate laminar patches in the near-wall flow, breaking spatial homogeneity in the wall-parallel directions.

\begin{figure}
\centering
\begin{tikzpicture}[node distance=0.4cm]

  \node(imgA) at (0,0) {\includegraphics[width=0.9\textwidth]{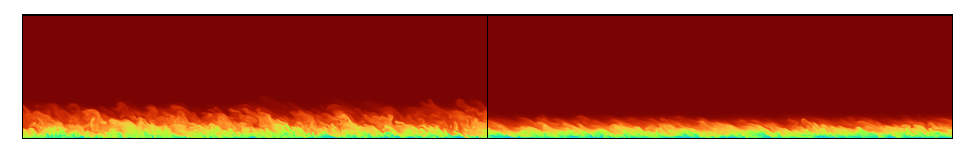}};
  \node[left=of imgA, yshift=1.25cm, anchor=north west] (labelA) {(a)};

  \node[below=of imgA]   (imgB) {\includegraphics[width=0.9\textwidth]{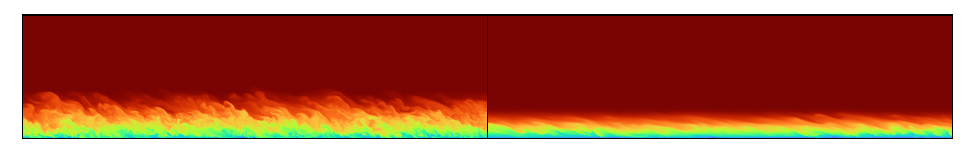}};
  \node[left=of imgB,yshift=1.25cm, anchor=north west] (labelB) {(b)};

  \node[below=of imgB]   (imgC) {\includegraphics[width=0.9\textwidth]{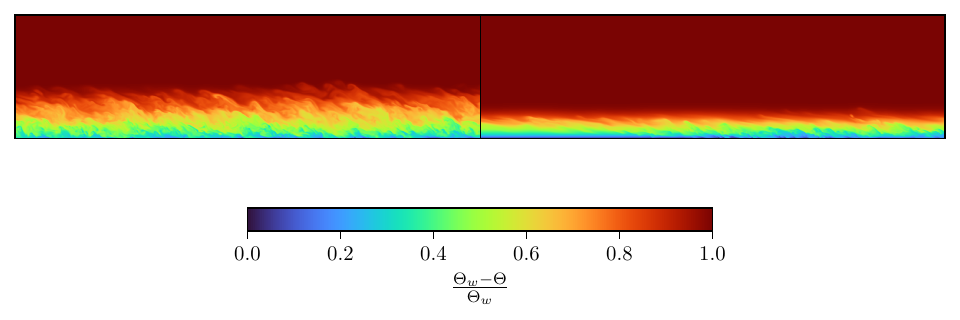}};
  \node[left=of imgC, yshift=2.45cm, anchor=north west] (labelC) {(c)};
\end{tikzpicture}
\caption{Temperature field contours in the plane $y=L_y/2$ in case B1 (left panels) and A3 (right panels) at times $Re_X = 30, 60$ and $90\times 10^5$ (identified by crosses in figure~\ref{fig:time_evolving_quantities}), from top to bottom. Spatial dimensions were scaled by $\nu/U_w$ to connect cases.}
\label{fig:tempfield_snapshots_yplane}
\end{figure}
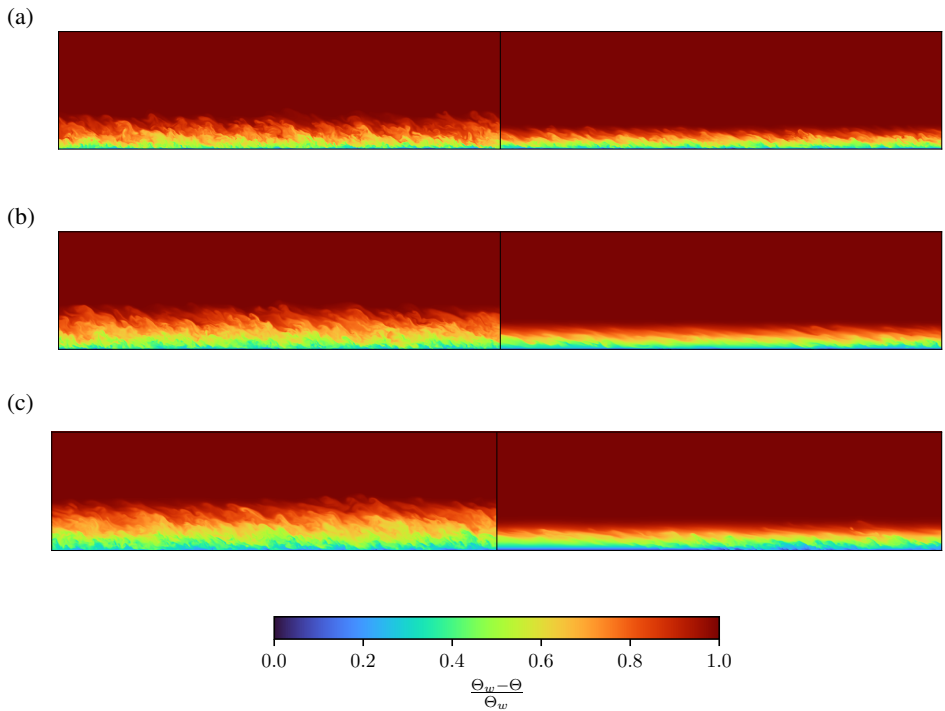

\begin{figure}
\centering
\includegraphics[width=.9\textwidth]{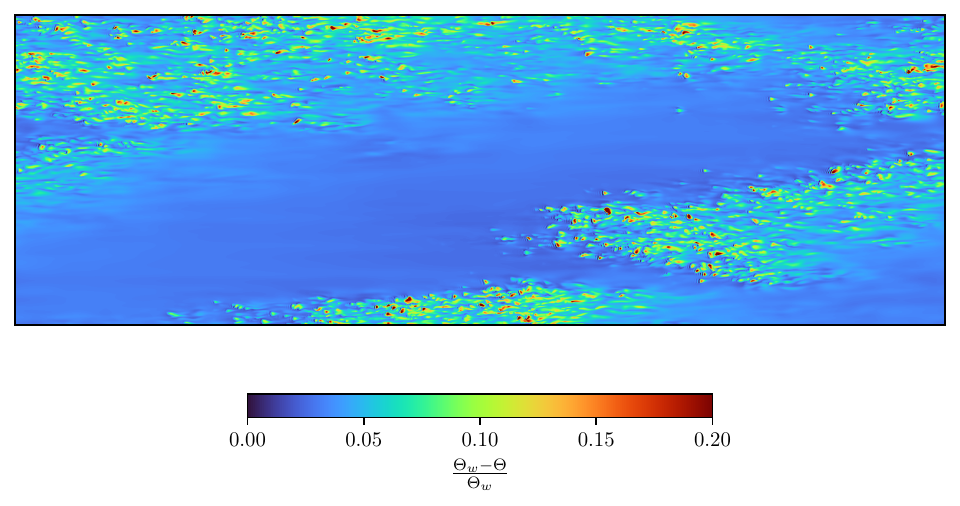}
\caption{Temperature field contour in the plane $z=10d_0$ (corresponding to $z^+ = 150$) in case A3 at time $Re_X = 120 \times 10^5$.}
\label{fig:tempfield_snapshot_zplane}
\end{figure}

Figure \ref{fig:tempfield_snapshots_yplane} shows snapshots of the temperature field over half of the streamwise extent of the domain at successive times (marked by black crosses in figure \ref{fig:time_evolving_quantities}), corresponding to $Re_X = 30 \times 10^5, \, 60\times 10^5$ and $90 \times 10^5$. This figure compares the two cases from table \ref{tab:parameters} that are the least  (B1, left panels) and the most (A3, right panels) affected by stratification, respectively. The temperature field in the left panels of figure \ref{fig:tempfield_snapshots_yplane} displays sustained turbulence driven by shear at the wall, although the growth of the boundary layer is inhibited by stratification compared to the neutral case. In case A3 (right panels), stratification effects act more rapidly, and turbulence is heavily damped, which significantly limits boundary-layer development. The snapshots also show spatial intermittency, with regions that are close to laminar while others still exhibit turbulent structures. The laminarization of the flow in case A3 is even more clearly observed in a plane parallel to the wall, as shown in figure \ref{fig:tempfield_snapshot_zplane}. While turbulence structures still subsist, very large portions of the flow are now laminar.

These observations can be connected to the time evolution of bulk quantities in figure \ref{fig:time_evolving_quantities} and to stability theories proposed in the literature. Different quantities have been proposed in the literature to predict the onset of turbulence collapse in stably stratified wall-bounded flows. From wind tunnel experiments, \cite{ohya1997} delineated the weakly stable and very stable regimes using a threshold bulk Richardson number of $Ri_\delta = 0.25$. These authors related this value to the results of the linear stability analysis of stratified shear flows by \cite{taylor1931} and \cite{miles1964}. This criterion was later debated in \cite{williams2017}, who reported the collapse of turbulence at different bulk Richardson numbers for smooth and rough walls, and for significantly lower values ($Ri_\delta \simeq 0.1-0.15$). This study argues that a better measure of regime transition is the wall-based gradient Richardson number, as originally proposed by \cite{schlichting1935}, with a critical value $Ri_{w,cr} = 1/24$ in the high-Reynolds-number limit. The ratio $\delta/L$, which is linked to the friction Richardson number $Ri_\tau$ \citep{flores2011}, has also been extensively used to predict the onset of relaminarization \citep{nieuwstadt_2005,ohya1997,vandewiel2007}. The values reported for the critical ratio $\left(\delta/L\right)_{c}$ above which turbulence starts to collapse vary widely between numerical studies, laboratory experiments and field data, with values ranging from $\left(\delta/L\right)_{c} = 0.14$ for \cite{coleman1992}, $\left(\delta/L\right)_{c}  = 1.2$ for \cite{nieuwstadt_2005}, to $\left(\delta/L\right)_{c}  = 5.0$ for \cite{ohya1997}. The possible Reynolds-number dependency of this criterion led \cite{flores2011} to introduce the Reynolds number based on the Obukhov length as an alternative criterion for the onset of turbulence collapse. Indeed, $Re_L$ represents the separation of scales between the distance from the wall at which buoyancy effects overcome shear production (capping the size of turbulent eddies) and the viscous length scale, thereby quantifying the extent of the (possible) dynamic sublayer. These authors argue that when $Re_L \lesssim 100$, separation of scales is not sufficient to have an inertial subrange, and connect this value to the commonly accepted $Re_\tau \simeq 100$ threshold below which turbulence collapses in a neutral channel flow. Figure \ref{fig:time_evolving_quantities}(c) shows that $Re_L$ falls below 100 around $Re_X \simeq 83 \times 10^5$ in case A3. Although the flow is not yet fully laminar, intermittent laminar--turbulent patches begin to appear around this value. In line with the findings of \cite{williams2017}, we observe in figure \ref{fig:time_evolving_quantities}(d) that turbulence intermittency occurs for values smaller than $Ri_\delta = 0.25$, as the final value reached by the bulk Richardson number in case A3 is 0.130. 

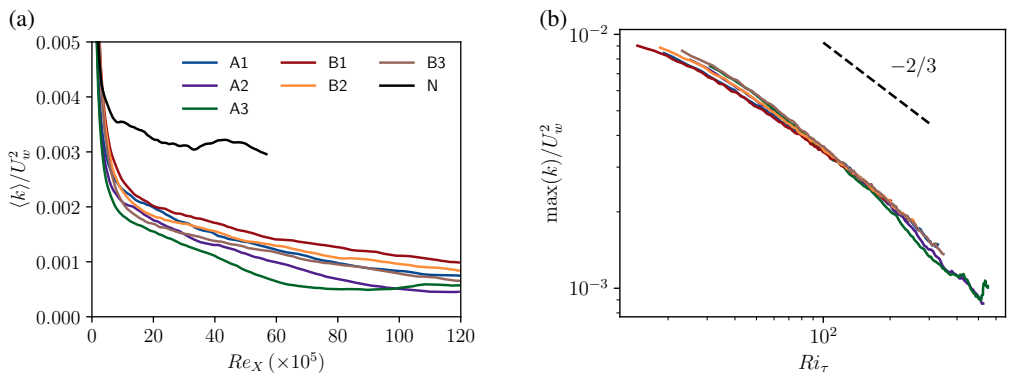
\begin{figure}
\begin{minipage}[b]{0.48\linewidth}
\includegraphics[width=\textwidth]{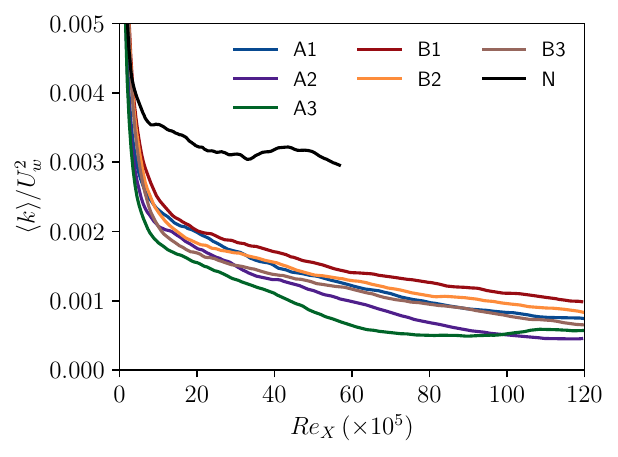}%
\begin{picture}(0,0)
    \put(-180,135){(a)}
\end{picture}
\end{minipage}
\hfill
\begin{minipage}[b]{0.48\linewidth}
\includegraphics[width=\textwidth]{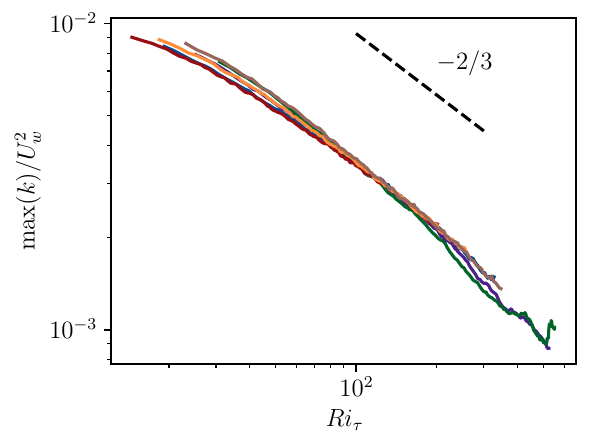}%
\begin{picture}(0,0)
    \put(-180,135){(b)}
\end{picture}
\end{minipage}
\caption{(a) Temporal evolution of the boundary-layer-averaged TKE. (b) Evolution of the maximum TKE across the boundary layer with $Ri_\tau$ in log-log-scale. Dashed line indicates $-2/3$ power law.}
\label{fig:TKE}
\end{figure}

Another way to characterize the influence of stratification on turbulence is through the temporal evolution of the turbulent kinetic energy (TKE) $k = 0.5\overline{u_i' u_i'}$, averaged over the height of the boundary layer, i.e. $\left<k\right> =(1/\delta)\smallint_0^\delta k(z) \, dz$. As observed in figure \ref{fig:TKE}(a), the boundary-layer-averaged TKE decreases faster in stably stratified cases with respect to the neutral case, and this decay is more pronounced for cases with lower $Re_0$ and larger $Ri_0$. In case A3, the boundary-layer-averaged TKE stops decaying and starts oscillating, marking the transition to global intermittency (or very stable regime) \citep{ansorge2014}. %
Figure \ref{fig:TKE}(b) shows the evolution of the maximum TKE across the boundary layer as a function of $Ri_\tau$ for the six stratified cases (A1--B3). Here, a marked collapse emerges at sufficiently high $Ri_\tau$ (after the transient regime) and before turbulence collapses, following an apparent $-2/3$ power law decay. The collapse suggests that the peak TKE magnitude is closely correlated with the friction Richardson number over this range.

\subsection{Similarity theory: global and local scaling}
\label{section:similarity_theory}
We now assess the fidelity of the similarity theories introduced in \S\ref{section:theory} using the present DNS dataset. Recall that, as mentioned in the introduction, this set-up has been successfully used to reproduce classical boundary-layer scaling results \citep{Kozul2016,cimarelli2024-1}. To understand its suitability for a stably stratified turbulent boundary layer, we examine the wall-normal profiles of first- and second-order statistics.
Since the boundary layer grows in time, successive instants of the simulation will correspond to different stratification levels, characterized by the ratio $\delta/L$. As the flow evolves, this ratio increases, and buoyancy constrains turbulence progressively closer to the wall relative to the boundary-layer thickness.

Based on field data in the weakly stable regime, different studies \citep{businger1971, dyer1974,brutsaert1982} proposed linear relations for the stability functions $\phi_m$ and $\phi_h$ as given by Eqs.~\eqref{eq:phi_m} and~\eqref{eq:phi_h}, 
where, in the neutral limit ($L \rightarrow \infty$, $\zeta \rightarrow 0$),  the Prandtl--von Kármán logarithmic velocity profile is recovered. Typical values $\beta_m = \beta_h \simeq 5.0$ are reported in the literature \citep{brutsaert1982,kaimal1994book}. For the intercept $\alpha_h$ -- which corresponds to the turbulent Prandtl number in neutral conditions $Pr_{t_0}$ -- values between $0.74$ \citep{businger1971} and $1.0$ \citep{kaimal1994book} have been proposed.  
Subsequent studies indicate that $\phi_m$ and $\phi_h$ deviate from a linear profile at higher stratification levels and become weaker functions of $\zeta$ for $\zeta > 1$. More complex expressions were proposed by \citet{cheng2005} and \citet{beljaars1991}, where $\phi_m$ and $\phi_h$ are approximately linear for $\zeta < 1 $ and level off for larger $\zeta$. It has also been suggested that the variability in the proposed empirical relations arises from the dependence of mean profiles on additional parameters, in particular the ABL depth $\delta$, or the Coriolis parameter $f$ \citep{zilitinkevich1989, heisel2023}. Since flow rotation is not included in this work, the present simulations provide an accurate assessment of the intrinsic effects of stratification on these scaling laws. Many rotating DNS and LES studies of the stably stratified ABL are driven by a geostrophic wind forcing, for which the boundary-layer depth scales with the Coriolis parameter $\delta \sim  u_\tau/f$ \citep{zilitinkevich1989}, as in \cite{coleman1992} and \cite{shah2014}. The stably stratified boundary layer studied in this work, in the absence of such a rotation-imposed depth constraint, instead continues to grow over time. Given its relevance to engineering flows with stable stratification, this canonical configuration provides a natural setting in which to assess the fidelity of MOST, as previously done for turbulent channel flows \citep{kotturshettar2025}.

\begin{figure}
\begin{minipage}[b]{0.48\textwidth}
    \centering
    \includegraphics[scale=.62]{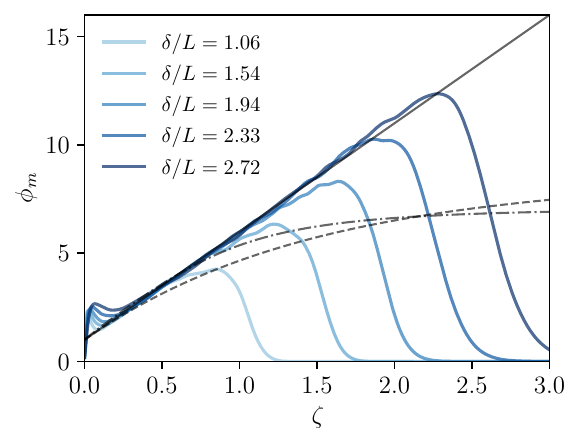}
    \begin{picture}(0,0)
        \put(-180,130){(a)}
    \end{picture}
    \label{fig:phi_m_a}
\end{minipage}
\hfill
\begin{minipage}[b]{0.48\textwidth}
    \centering
    \includegraphics[scale=.62]{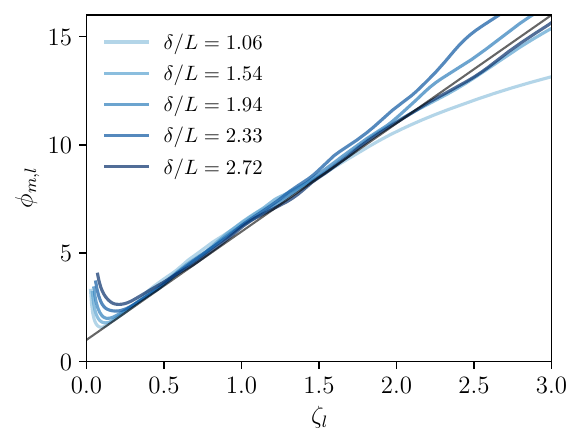}
    \put(-180,130){(b)}
    \label{fig:phi_m_b}
\end{minipage}

\begin{minipage}[b]{0.48\textwidth}
    \centering
    \includegraphics[scale=.62]{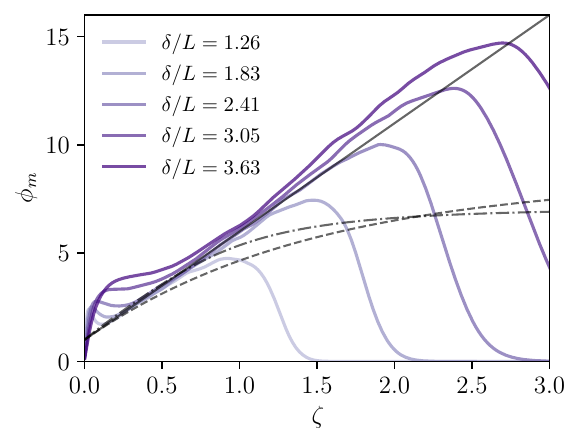}
    \put(-180,130){(c)}
    \label{fig:phi_m_c}
\end{minipage}
\hfill
\begin{minipage}[b]{0.48\textwidth}
    \centering
    \includegraphics[scale=.62]{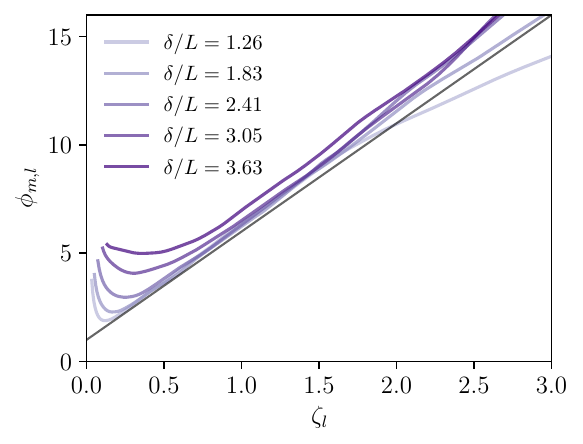}
    \put(-180,130){(d)}
    \label{fig:phi_m_d}
\end{minipage}

\begin{minipage}[b]{0.48\textwidth}
    \centering
    \includegraphics[scale=.62]{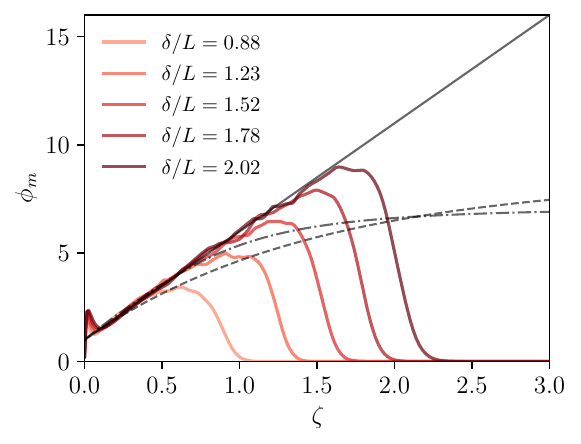}
    \put(-180,130){(e)}
    \label{fig:phi_m_e}
\end{minipage}
\hfill
\begin{minipage}[b]{0.48\textwidth}
    \centering
    \includegraphics[scale=.62]{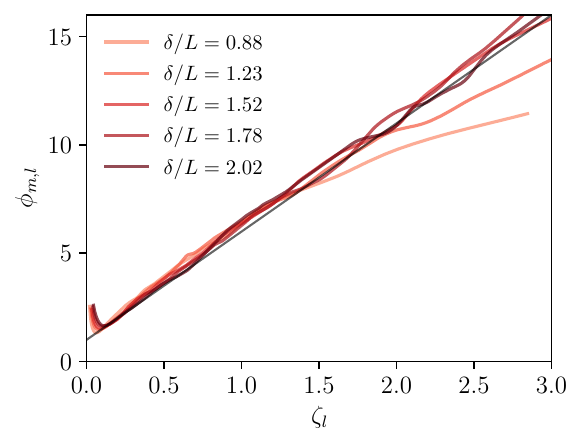}
    \put(-180,130){(f)}
    \label{fig:phi_m_f}
\end{minipage}
\caption{Mean velocity gradient in global scaling ((a), (c), (e)) and local scaling ((b), (d), (f)) at different instants of the flow in case A1 ((a), (b)), A2 ((c), (d)) and B1 ((e), (f)). DNS results are compared with the similarity relation $\phi_m = 1 + 5.0 \zeta$ \citep{dyer1974,brutsaert1982} (black solid line) and non-linear relations proposed by \cite{beljaars1991} (black dashed line) and \cite{cheng2005} (black dash-dotted line) in the global panels. In the local scaling panels, the linear relation is $\phi_{m,l}=1+5.0\zeta_l$.}
\label{fig:phi_m}
\end{figure}

In this study, we will focus on the scaling of velocity field statistics. Additional details on the mean temperature profile can be found in Appendix \ref{appendix:temperature}. 
The wall-normal profiles for the similarity function $\phi_m$ obtained from DNS are presented in figure~\ref{fig:phi_m}(a), (c) and (e) for the three cases A1, A2 and B1, at several instants (corresponding to increasing levels of stratification) and in global scaling. %
First, it can be observed that the linear scaling given by Eq. \eqref{eq:phi_m} (with slope $\beta_m = 5.0$) is well satisfied for the three cases over a wide range of stratification levels, and beyond the approximately constant-flux region and the range $\zeta \lesssim 0.1$ associated with surface scaling in the SHEBA observations \citep{grachev2005}. Only for case A2 (figure \ref{fig:phi_m}(c)) at later times (i.e. large $\delta/L$), the scaled velocity gradient seems to depart from the linear similarity relation. %
\cite{shah2014} performed a DNS of a stably stratified TBL with geostrophic wind forcing and reported a slope $\beta_m \simeq 9.0$ -- an unusually large value they attributed to a Reynolds-number dependence of the similarity theory, based on simulations spanning a similar range of friction Reynolds numbers ($Re_\tau$ $ < 1200$) to that in the present work. Our DNS, however, reaches even lower Reynolds numbers ($Re_\tau < 830$ in stratified cases; see table~\ref{tab:parameters}) and yet recovers the classical value $\beta_m \simeq 5.0$. While Reynolds-number effects on this scaling cannot be entirely ruled out, this suggests that other mechanisms -- such as the Coriolis effect, absent from the present configuration -- may be responsible for the larger $\beta_m$ reported by \cite{shah2014}. 

Considering now the departure from linear scaling at large stratification levels, we note that $\phi_m$ levels off in figure \ref{fig:phi_m}(a) before collapsing, but the level of this plateau seems related to the proximity of the boundary-layer edge ($z$ approaching $\delta$) rather than to the value of $\zeta$. Consequently, non-linear relations proposed by \citet{cheng2005} or \citet{beljaars1991} do not match our DNS profiles since they do not incorporate $\delta$ in their expression. Finally, we attempted to account for the effect of the boundary-layer thickness using the mixed scaling parameter $Z = \sqrt{\delta L}$ that was recently introduced by \cite{heisel2023}. %
No successful collapse could be obtained across our DNS data after scaling the wall distance $z$ by this new parameter. However, this scaling was proposed from an LES study of a rotating stably stratified ABL, where $\delta$ reflects a rotation-imposed equilibrium (as discussed above), which may explain why the mixed parameter $Z$ does not collapse our data in the same way. 

Figures \ref{fig:phi_m}(b), (d) and (f) reproduce the analysis of figures \ref{fig:phi_m}(a), (c) and (e) but in {\it local} scaling, i.e. using the local Obukhov length and local shear stress to scale the velocity gradient.
In this case, the plots have been arbitrarily started from $z=0.02 \delta$ since the velocity gradient in local scaling diverges very close to the wall. 
The agreement with the linear similarity theory is also observed across all three cases over a wide range of $\zeta_l$ values. Since the local Obukhov length $\Lambda$ decreases in the outer layer, points farther from the surface correspond to large $\zeta_l$. As a consequence, the outer layer appears artificially expanded in $\zeta_l$-space when using local scaling.
Based on figure \ref{fig:phi_m}, global and local scaling theories exhibit similar performance over a wide range of $\zeta$ (and $\zeta_l$) values. This extends the range of validity of the surface scaling relative to the SHEBA observations of \cite{grachev2005}, in which departures from the approximately constant-flux regime occur for $\zeta \gtrsim 0.1$ and local scaling performs better. %

\begin{figure}
\begin{minipage}[b]{0.48\linewidth}
\includegraphics[scale=0.6]{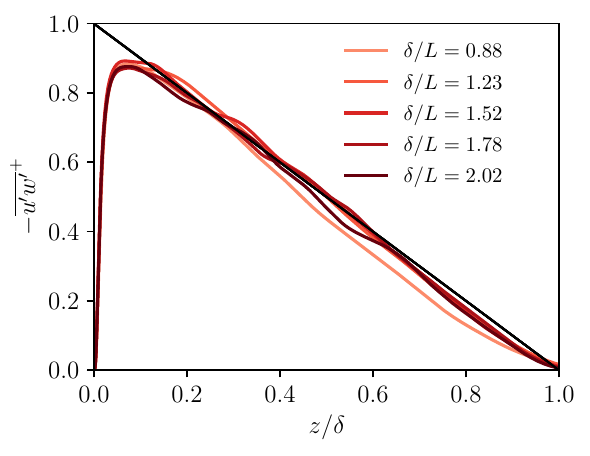}%
\begin{picture}(0,0)
    \put(-170,125){(a)}
\end{picture}
\end{minipage}
\hfill
\begin{minipage}[b]{0.48\linewidth}
\includegraphics[scale=0.6]{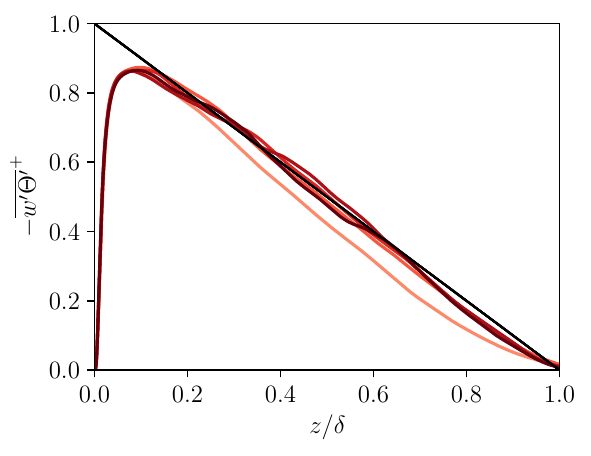}%
\begin{picture}(0,0)
    \put(-170,125){(b)}
\end{picture}
\end{minipage}
\caption{Wall-normal profiles of (a) turbulent shear stress and (b) turbulent heat flux in case B1. Solid black line indicates the linear decay $\tau/u_\tau^2 = 1 - z/\delta$ in (a) and $q/q_w = 1 - z/\delta$ in (b).}
\label{fig:turbulent_fluxes}
\end{figure}

Figure~\ref{fig:turbulent_fluxes} presents the wall-normal profiles of the turbulent shear stress and turbulent heat flux, for case B1, showing that both decay approximately linearly with $z/\delta$ far from the surface, as in neutral conditions\footnote{For comparison, the steady-state model of \cite{nieuwstadt1984} gives $q/q_w = 1-z/h$ and $\tau/u_\tau^2 = (1-z/h)^{3/2}$, where $h$ is the boundary-layer height.}; i.e. $\tau/u_\tau^2 \sim 1-z/\delta$ and $q/q_w \sim 1-z/\delta$. The same trend is observed for the five other cases in table \ref{tab:parameters} (not shown here). With these results in mind, let us consider the $z$-less regime ($z/L \rightarrow \infty$), where the momentum similarity function in local scaling can be approximated by $\phi_{m,l} \simeq \beta_m \zeta_l$. %
Assuming that the fluxes $\tau$ and $q$ share the same linear decay (as supported by figure~\ref{fig:turbulent_fluxes}), it follows that the mean velocity gradient can be expressed as
\begin{equation}
    \frac{\partial \bar u}{\partial z} \simeq \beta_m \frac{u_*}{\kappa \Lambda} = -\beta_m \frac{\beta g q(z) }{\tau(z)} = -\beta_m \frac{\beta g q_w }{u_\tau^2}.
    \label{eq:zless_scaling}
\end{equation}
Eq.~\eqref{eq:zless_scaling} implies that global and local scaling theories become equivalent for the mean velocity gradient in the $z$-less limit. %

\begin{figure}
    \centering
    \includegraphics[width=0.49\linewidth]{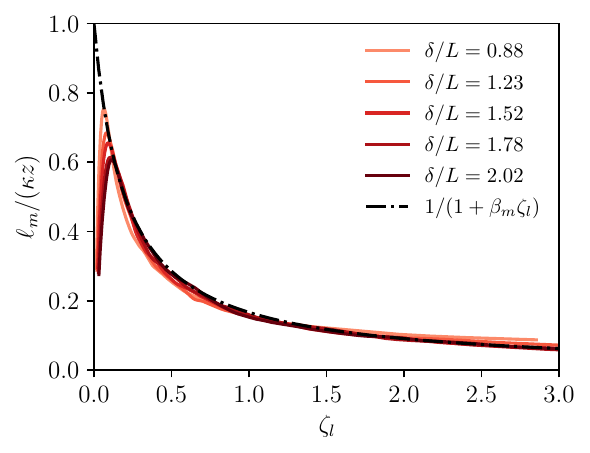}
    \caption{Wall-normal profile of Prandtl's mixing length in local scaling in case B1. The black dash-dotted curve corresponds to Eq.~\eqref{eq:mixing_length}.}
    \label{fig:mixing_length}
\end{figure}

As both global and local scaling approaches yield a similar form for the mean velocity gradient in the $z$-less limit, it is worth connecting MOST to Prandtl's mixing-length theory for further insight. The eddy viscosity model for the turbulent shear stress is given by 
\begin{equation}
    -\overline{u'w'} = u_*^2 = K_m \frac{\partial \bar u}{\partial z} = \ell_m^2 \left(\frac{\partial \bar u}{\partial z} \right)^2,
\end{equation}
where $K_m$ is the turbulent viscosity and $\ell_m$ is Prandtl’s mixing length. Then, it follows from MOST in local scaling that $\ell_m$ can be expressed as 
\begin{equation}
    \ell_m = \kappa z \phi_{m,l}^{-1} = \frac{\kappa z}{1+ \beta_m \frac{z}{\Lambda}}. 
    \label{eq:mixing_length}
\end{equation}
This has the same algebraic form as the classical Blackadar mixing-length expression for the planetary boundary layer \citep{blackadar1962}: 
\begin{equation}
    \ell_m  = \frac{\kappa z}{1+ \frac{\kappa z}{\ell_\infty}}, 
\end{equation}
where $\ell_\infty$ corresponds here to the local limiting scale $\kappa \Lambda(z)/\beta_m$. %
Consistent with the previous observations, figure \ref{fig:mixing_length} confirms that the collapse between the theoretical model for $\ell_m$ (Eq. \eqref{eq:mixing_length}) and its direct evaluation from the local shear stress profile obtained from DNS is excellent. All other stratified cases show similar agreement (not shown here). 

\begin{figure}
\begin{minipage}[b]{0.48\linewidth}
\includegraphics[width=.96\linewidth]{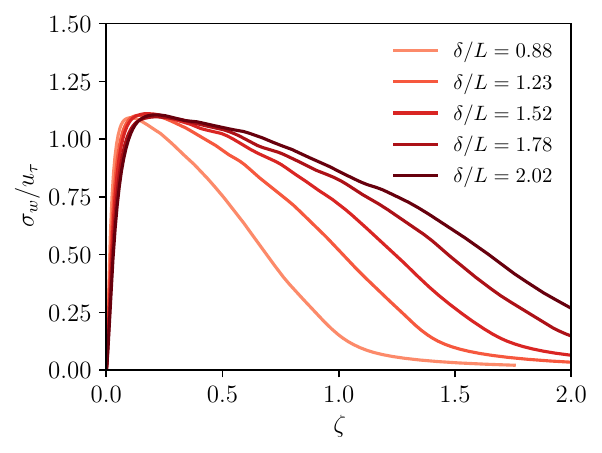}%
\begin{picture}(0,0)
    \put(-175,130){(a)}
\end{picture}
\end{minipage}
\hfill
\begin{minipage}[b]{0.47\linewidth}
\includegraphics[width=.96\linewidth]{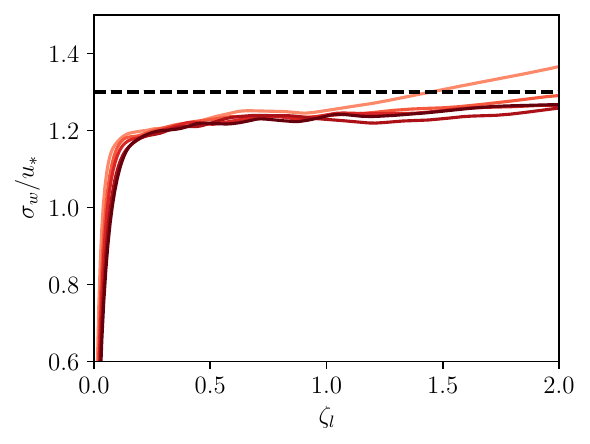}%
\begin{picture}(0,0)
    \put(-175,130){(b)}
\end{picture}
\end{minipage}
\caption{Wall-normal profile of vertical velocity root mean square in (a) global scaling and (b) local scaling, in case B1. Dashed line in (b) indicates the asymptotic value $\sigma_w/u_*=1.3$ from \cite{grachev2015}.}
\label{fig:sigma_w}
\end{figure}
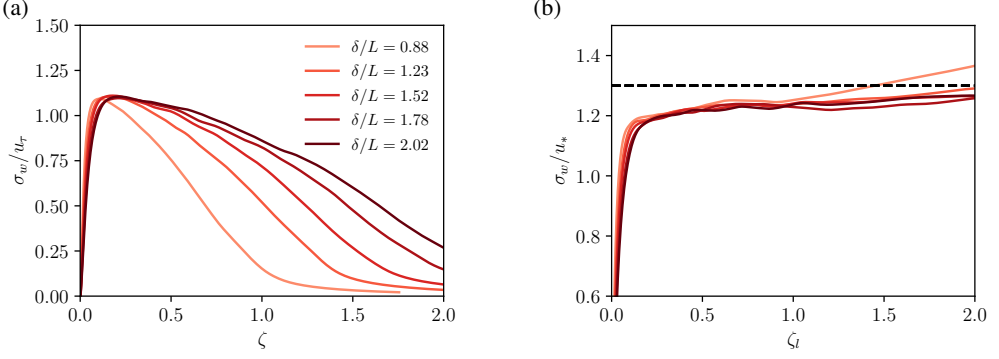

Finally, MOST can be used to predict the scaling of higher-order turbulence statistics such as the root mean square (rms) of the wall‐normal velocity component $ {\sigma_w = \sqrt{\overline{w'^2}}}$; see \cite{nieuwstadt1984}. In figure~\ref{fig:sigma_w}, we compare the wall-normal profiles of $\sigma_w$ in global and local scaling for case B1. Unlike for the mean velocity gradient, the global scaling theory cannot provide a universal profile for $\sigma_w$ across a wide range of stratification levels. Conversely, the locally scaled rms profiles collapse reasonably well on a single curve and approach an asymptotic value consistent with the results of \cite{grachev2015}: $\sigma_w/u_* = \phi_w = 1.3$. Deviations from the asymptotic limit near the edge of the boundary layer may reflect limited statistical convergence of the DNS data in this region of the flow, yet the analysis still supports the validity of local scaling theory across a wide range of stratification levels. 

Hence, the results presented in this section indicate that, for a stably stratified non-rotating TBL where turbulence is controlled only by the wall-driven shear–buoyancy balance, scaling of the mean velocity gradient with surface fluxes remains valid over a wider $\zeta$-range than previously reported, provided that turbulent fluxes decay with height at the same rate. However, the profiles of the mixing length and wall-normal velocity root mean square confirm that \citeauthor{nieuwstadt1984}'s local scaling theory is a more robust modelling framework.

\subsection{Mean velocity profile} 

\begin{figure}
\begin{minipage}[b]{0.48\textwidth}
    \centering
    \includegraphics[scale=.64]{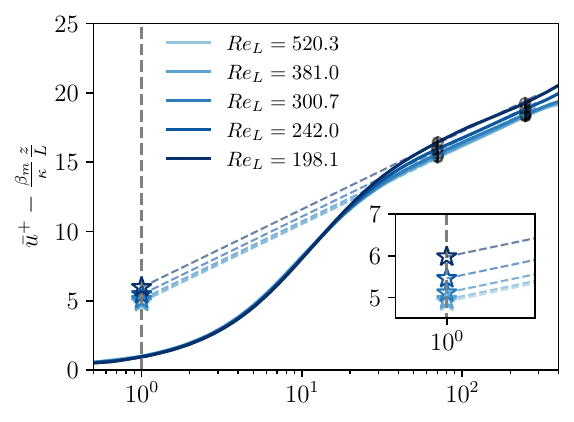}
    \begin{picture}(0,0)
        \put(-180,125){(a)}
    \end{picture}
    \label{fig:mean_velocity_profile_a}
\end{minipage}
\hfill
\begin{minipage}[b]{0.48\textwidth}
    \centering
    \includegraphics[scale=.64]{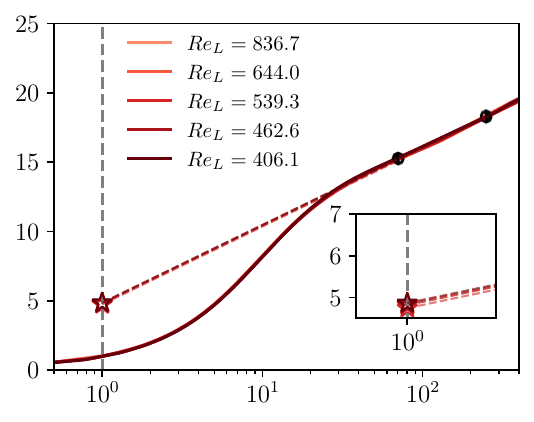}
    \begin{picture}(0,0)
        \put(-180,125){(b)}
    \end{picture}
    \label{fig:mean_velocity_profile_b}
\end{minipage}

\begin{minipage}[b]{0.48\textwidth}
    \centering
    \includegraphics[scale=.64]{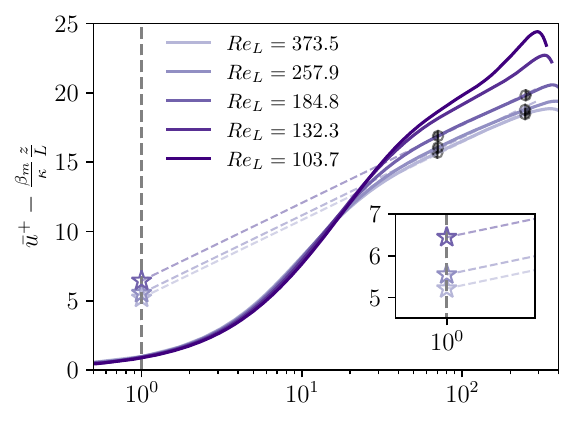}
    \begin{picture}(0,0)
        \put(-180,125){(c)}
    \end{picture}
    \label{fig:mean_velocity_profile_c}
\end{minipage}
\hfill
\begin{minipage}[b]{0.48\textwidth}
    \centering
    \includegraphics[scale=.64]{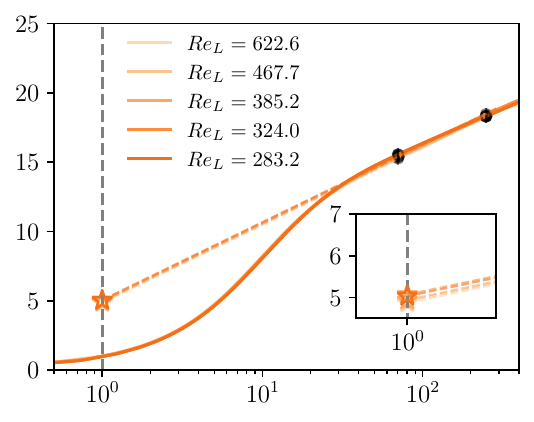}
    \begin{picture}(0,0)
        \put(-180,125){(d)}
    \end{picture}
    \label{fig:mean_velocity_profile_d}
\end{minipage}

\begin{minipage}[b]{0.48\textwidth}
    \centering
    \includegraphics[scale=.64]{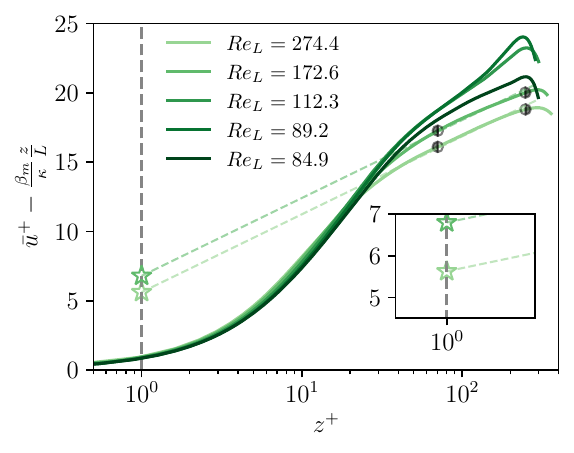}
    \begin{picture}(0,0)
        \put(-180,135){(e)}
    \end{picture}
    \label{fig:mean_velocity_profile_e}
\end{minipage}
\hfill
\begin{minipage}[b]{0.48\textwidth}
    \centering
    \includegraphics[scale=.64]{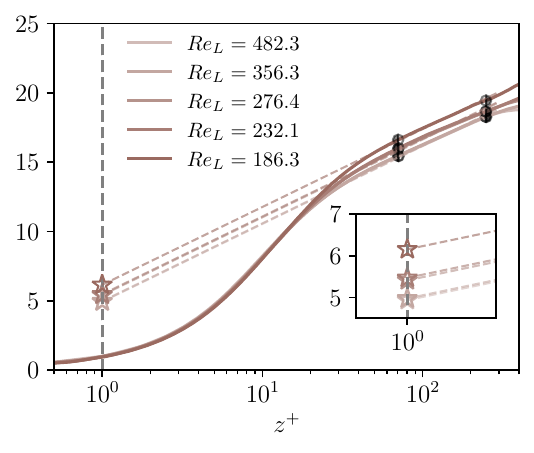}
    \begin{picture}(0,0)
        \put(-180,135){(f)}
    \end{picture}
    \label{fig:mean_velocity_profile_f}
\end{minipage}
\caption{Wall-normal profiles of inner-scaled mean streamwise velocity component with  the Monin--Obukhov linear contribution subtracted in case A1 to A3 ((a), (c), (e)) and B1 to B3 ((b), (d), (f)). The log-law intercept has been fitted by regression in the interval $z^+ \in [70, 250]$ (indicated by half-filled circles) for cases with $Re_L > 150$; the fit is highlighted in the insets.}
\label{fig:mean_velocity_profile}
\end{figure}
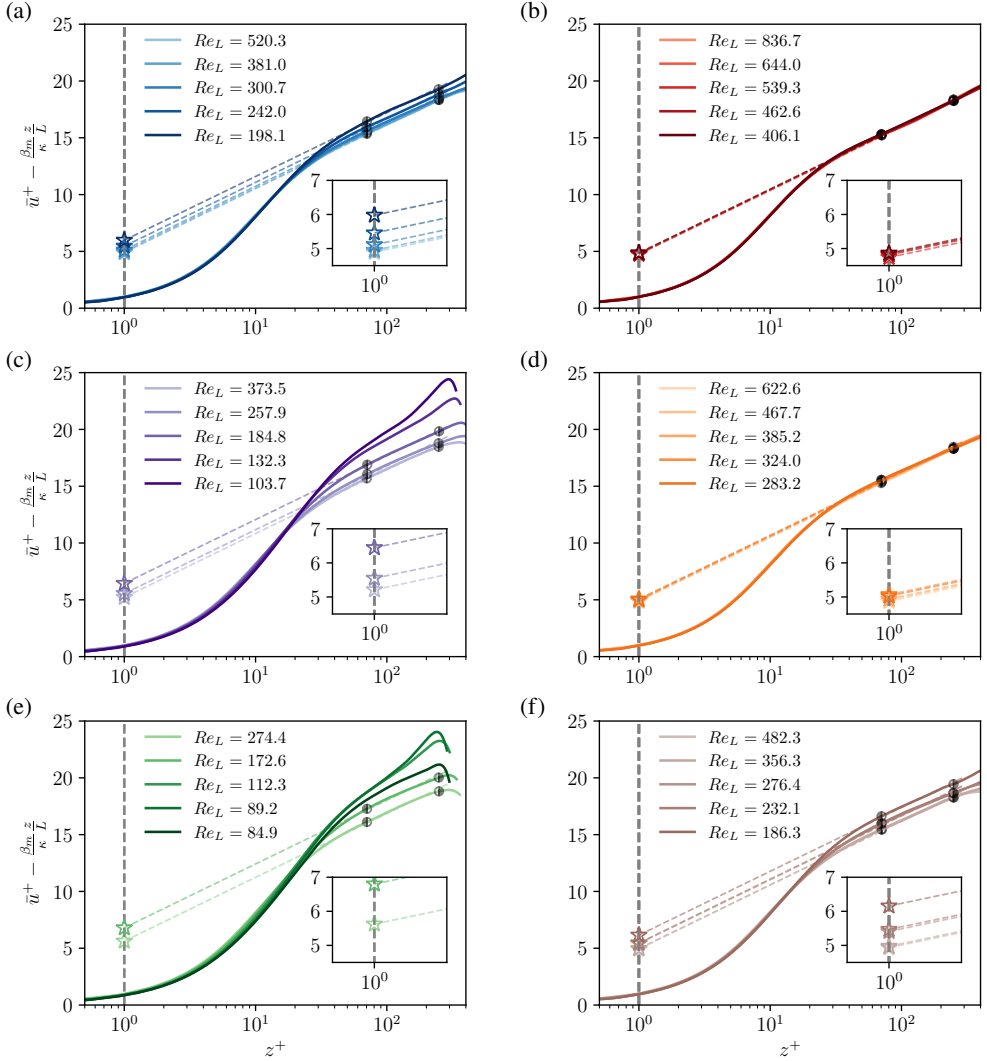

\begin{figure}
\begin{minipage}[b]{0.48\textwidth}
    \centering
    \includegraphics[scale=.64]{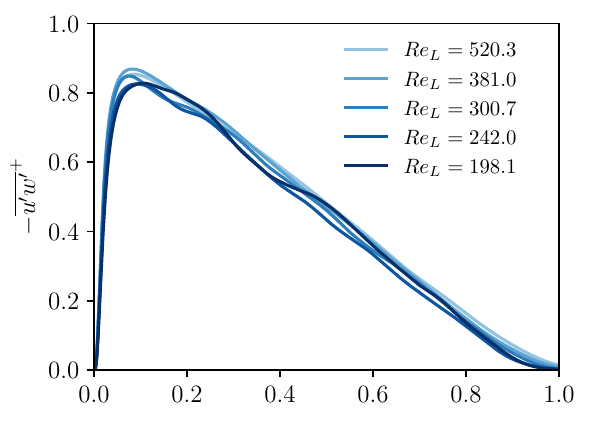}
    \begin{picture}(0,0)
        \put(-180,125){(a)}
    \end{picture}
\end{minipage}
\hfill
\begin{minipage}[b]{0.48\textwidth}
    \centering
    \includegraphics[scale=.64]{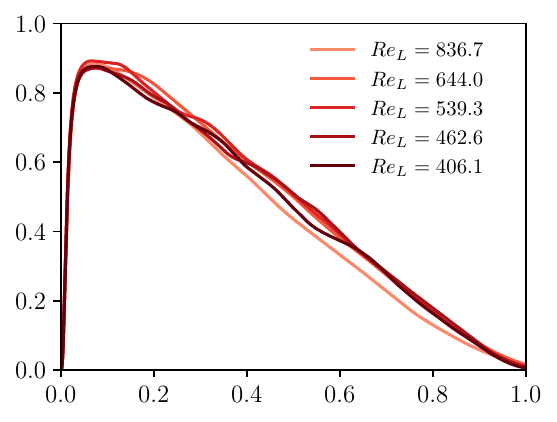}
    \begin{picture}(0,0)
        \put(-180,125){(b)}
    \end{picture}
\end{minipage}

\begin{minipage}[b]{0.48\textwidth}
    \centering
    \includegraphics[scale=.64]{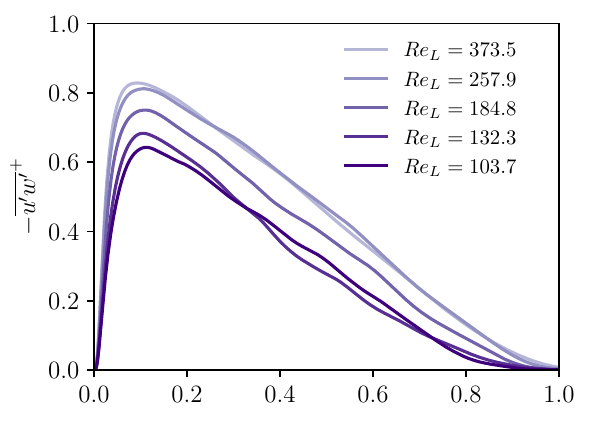}
    \begin{picture}(0,0)
        \put(-180,125){(c)}
    \end{picture}
\end{minipage}
\hfill
\begin{minipage}[b]{0.48\textwidth}
    \centering
    \includegraphics[scale=.64]{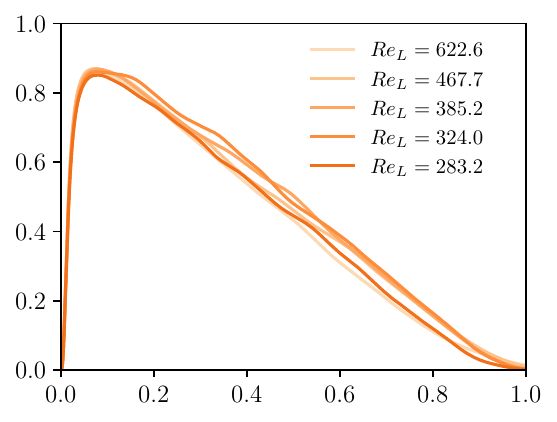}
    \begin{picture}(0,0)
        \put(-180,125){(d)}
    \end{picture}
\end{minipage}

\begin{minipage}[b]{0.48\textwidth}
    \centering
    \includegraphics[scale=.64]{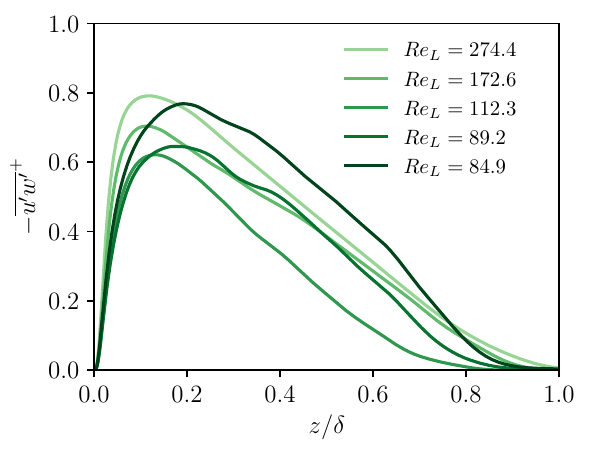}
    \begin{picture}(0,0)
        \put(-180,135){(e)}
    \end{picture}
\end{minipage}
\hfill
\begin{minipage}[b]{0.48\textwidth}
    \centering
    \includegraphics[scale=.64]{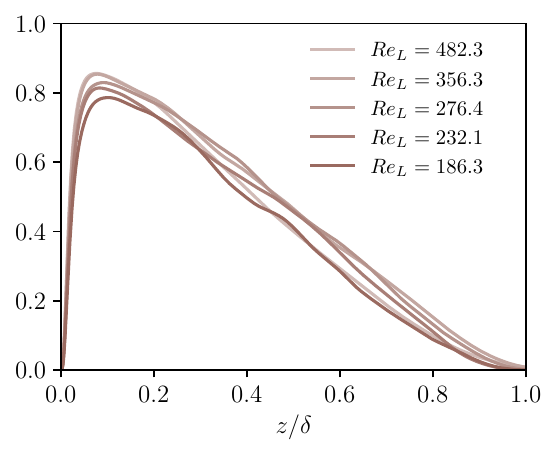}
    \begin{picture}(0,0)
        \put(-180,135){(f)}
    \end{picture}
\end{minipage}
\caption{Wall-normal profile of inner-scaled turbulent shear stress in cases A1 to A3 ((a), (c), (e)) and B1 to B3 ((b), (d), (f)).}
\label{fig:shear_stress_profile}
\end{figure}
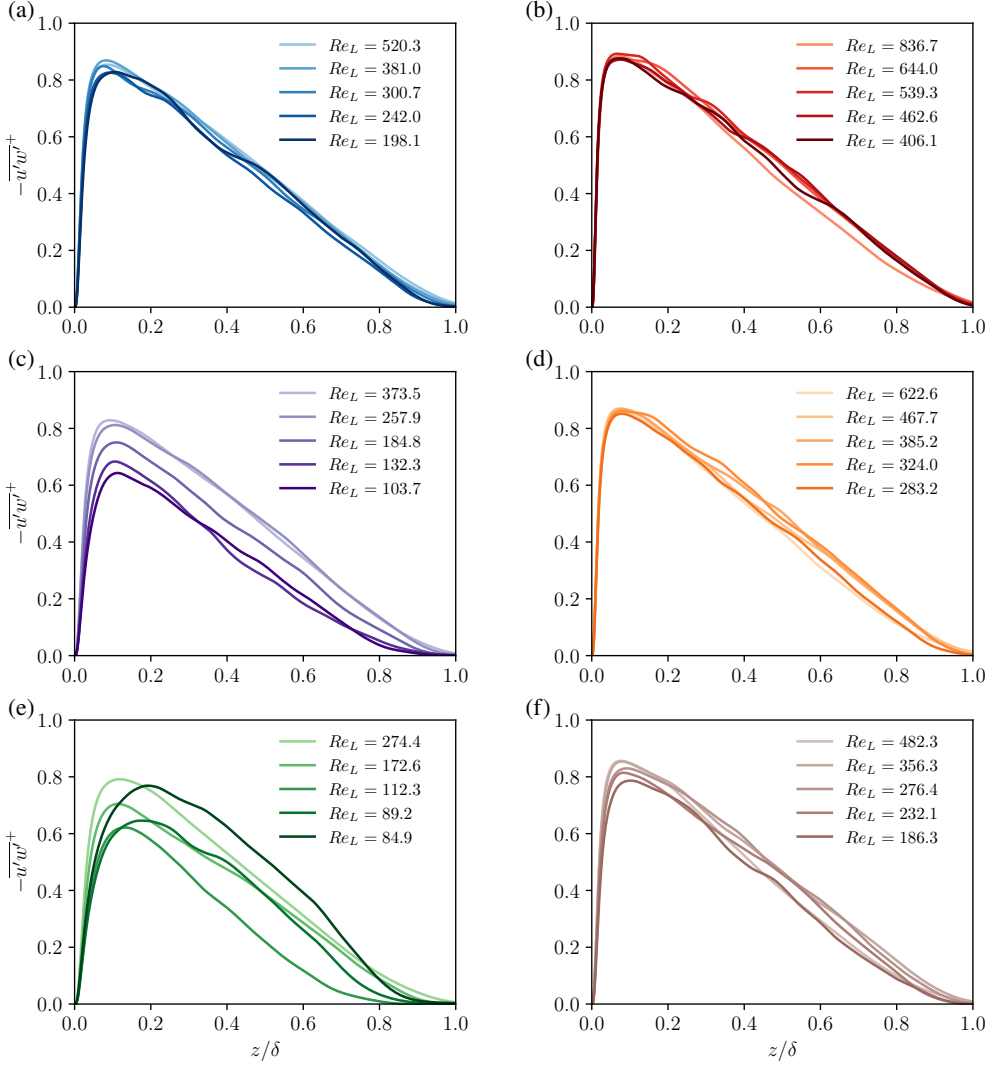

As we have seen, the mean velocity gradient in global scaling follows a linear trend across a wide range of stratification levels. %
Integration of Eq.~\eqref{eq:MOST} along $z$, combined with the expressions of $\phi_m$ %
given by Eq.~\eqref{eq:phi_m},  %
yields the following log-linear velocity profile:
\begin{align}
     \bar u^+ & = \frac{1}{\kappa}\log \left(z^+\right) + \Psi_m(z) + A, 
     \label{eq:loglinear_profile_vel} 
\end{align}
where $\Psi_m (z) = \beta_m z /(\kappa L)$ is the linear contribution arising from MOST with the Businger--Dyer linear closure for $\phi_m$, and $A$ is the profile intercept. Note that the contribution of $\Psi_m\left(z_0\right) = \beta_m/(\kappa Re_L)$, where $z_0 = \nu/u_\tau$, has been neglected as it  usually remains very small with respect to $\Psi_m(z)$, even for large stratification levels (see figure~\ref{fig:time_evolving_quantities}(c)). In figure~\ref{fig:mean_velocity_profile}, we report the inner-scaled mean streamwise velocity profile, compensated for the Monin--Obukhov contribution, for the six different cases A1 to B3 and for five successive times, here identified by the corresponding $Re_L$ value. For the cases that are least impacted by stratification (B1, B2), $Re_L$ remains large throughout the simulation, and we recover a classical log-law profile after subtraction of the linear component, with an intercept $A\simeq 4.9$ in line with the neutral-case value. However, as stratification intensifies, $Re_L$ decreases, leading to an upward shift in the log-linear profile. For $Re_L$ in an intermediate range ($400 \gtrsim Re_L \gtrsim 150 $), a logarithmic region with constant slope $1/\kappa \simeq 2.44$ still exists, but the intercept of the velocity profile increases with respect to the neutral-case value. However, \cite{armenio2002a} instead report an \textit{increase} in the slope and a \textit{decrease} in the additive constant of the logarithmic profile in their LES study of a stably stratified channel flow. \cite{atoufi2019} came to the same conclusion from DNS of a turbulent boundary layer developing over a cooled surface. However, both analyses are based on the inner-scaled mean velocity profile, without subtracting the linear component arising from the similarity theory. %
The upward shift in the intercept that we observe can actually be linked to a decrease in the peak value of the inner-scaled turbulent shear stress, as shown in figures \ref{fig:shear_stress_profile}(c) and \ref{fig:shear_stress_profile}(e). Indeed, as turbulent shear stress decreases relative to viscous forces, the viscous sublayer extends farther from the wall, and the log-law region shifts upward. %

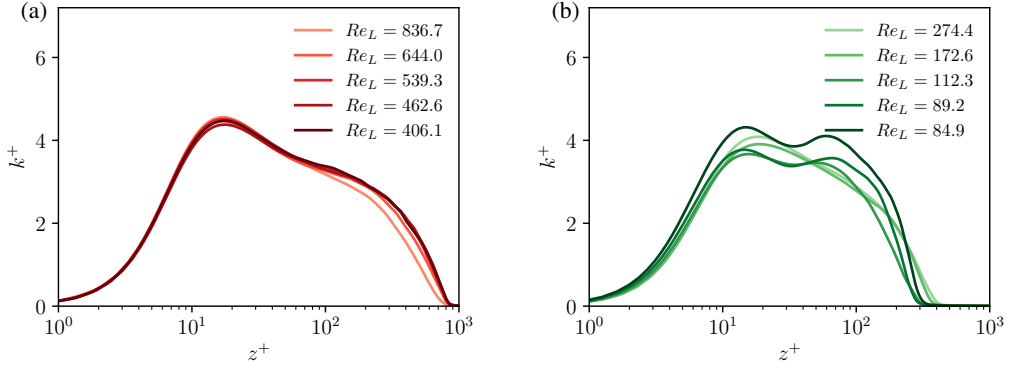
\begin{figure}
\begin{minipage}[b]{0.48\linewidth}
\includegraphics[width=\textwidth]{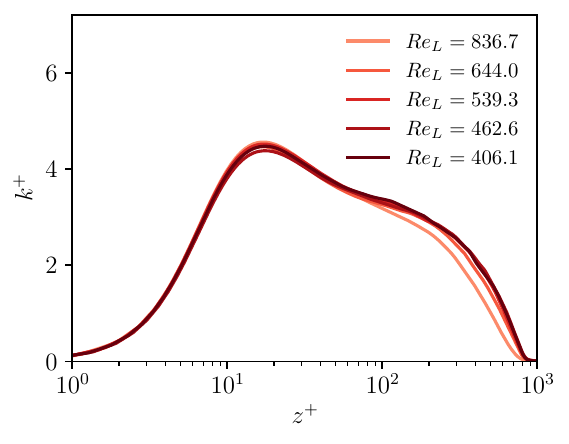}%
\begin{picture}(0,0)
    \put(-175,135){(a)}
\end{picture}
\end{minipage}
\hfill
\begin{minipage}[b]{0.48\linewidth}
\includegraphics[width=\textwidth]{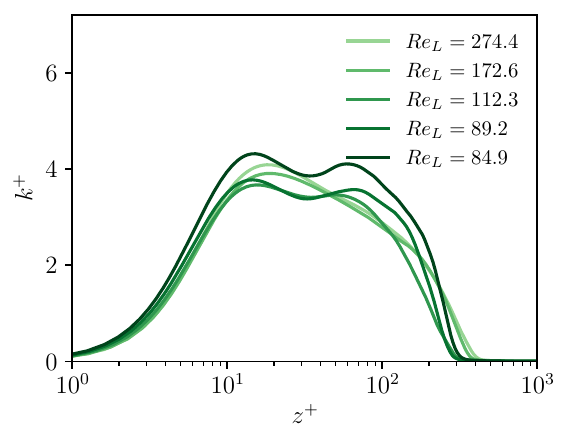}%
\begin{picture}(0,0)
    \put(-175,135){(b)}
\end{picture}
\end{minipage}
\caption{Wall-normal profiles of inner-scaled turbulent kinetic energy in cases B1 (a) and A3 (b).}
\label{fig:TKE_profile}
\end{figure}

Then, when $Re_L$ falls below a threshold -- here estimated to be around $150$ -- the separation of scales between the Obukhov length (which constrains the scale of the largest eddies in the dynamic sublayer \citep{flores2011}) and the viscous length scale $\nu/u_\tau$ is insufficient to allow for the existence of a logarithmic region. Hence, for $Re_L < 150$, fitting the log-law intercept becomes ill-posed. These observations are consistent with the $Re_L \approx 100$ threshold of \cite{flores2011} for the collapse of turbulence in a nocturnal boundary layer, and with the study of \cite{williams2017}, who observed turbulence collapse in their experiments for $Re_L \lesssim 130 \pm 60$. It can also be observed in figure \ref{fig:shear_stress_profile}(e) that, for case A3, when $Re_L$ further decreases, the location of the shear stress peak shifts towards the outer layer while its height increases in inner scaling ($u_\tau$ rapidly decreases when relaminarization sets in while velocity fluctuations oscillate, as discussed above). Another symptom of the breakdown of a self-sustained turbulent flow in case A3 can be seen in the wall-normal profiles of TKE shown in figure~\ref{fig:TKE_profile}(b). When the flow can no longer accommodate a dynamic sublayer ($Re_L$ being too low) and starts transitioning, the inner-scaled TKE profile is  damped (in contrast to case B1 shown in figure \ref{fig:TKE_profile}(a)), and a second peak emerges in the outer layer, stemming from the intermittent laminar--turbulent patches featured in the flow\footnote{Note that the nature of this second peak is different from the one predicted to be featured in a turbulent boundary layer at high Reynolds number (described, e.g. in \cite{marusic2010high}).}. 

\begin{figure}
    \centering
    \includegraphics[width=0.5\linewidth]{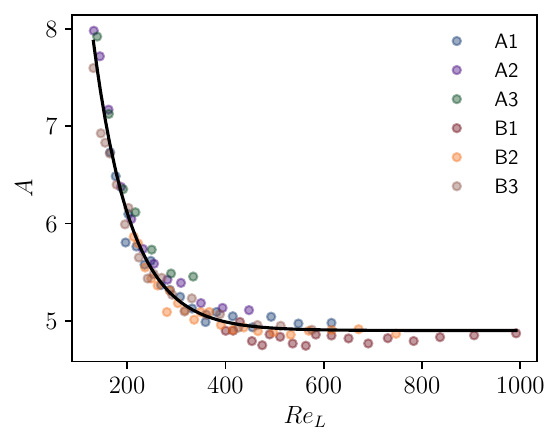}
    \caption{Evolution of the log-linear velocity profile intercept with $Re_L$ for the six stratified cases. The solid black line corresponds to the regression given by Eq. \eqref{eq:A_fit_ReL}.}
    \label{fig:log_law_intercept_vs_ReL}
\end{figure}

The intercept $A$ of the log-linear velocity profile defined by Eq.~\eqref{eq:loglinear_profile_vel} has been fitted for the six cases A1 to B3 and for successive times, as long as the criterion $Re_L > 150$ was met and in the interval $z^+ \in [70, 250]$ \footnote{A $Re_\tau$-dependent upper bound would be a more systematic choice given the range of friction Reynolds numbers across cases; however, a visual inspection of Figure \ref{fig:mean_velocity_profile} confirms that the chosen $z^+$ interval falls within the log-law region for all cases considered here.} . Below that threshold, the regression was no longer conducted. The values obtained for $A$ have been reported for the six cases as a function of $Re_L$ in figure \ref{fig:log_law_intercept_vs_ReL}. Remarkably, the data points collapse well on a master curve described by 
\begin{equation}
    A = 16.0 \exp(-0.0128 Re_L) + 4.9,
    \label{eq:A_fit_ReL} 
\end{equation}
with little sensitivity to the prescribed nominal Reynolds and Richardson numbers (the coefficient of determination is $R^2 = 0.97$). In line with our previous findings, the intercept remains constant and approximately equal to the neutral-case value ($A\simeq 4.9$) for $Re_L > 400$. Then, as $Re_L$ decreases, the intercept $A$ displays exponential growth. It can be noted that, for some cases, the data points corresponding to the largest $Re_L$ values are slightly off-trend. This must be attributed to the still ongoing destabilization of the shear layer and the temporal boundary layer not yet having completed its initial transient. 

\begin{figure}
\begin{minipage}[b]{0.48\textwidth}
    \centering
    \includegraphics[width=\linewidth]{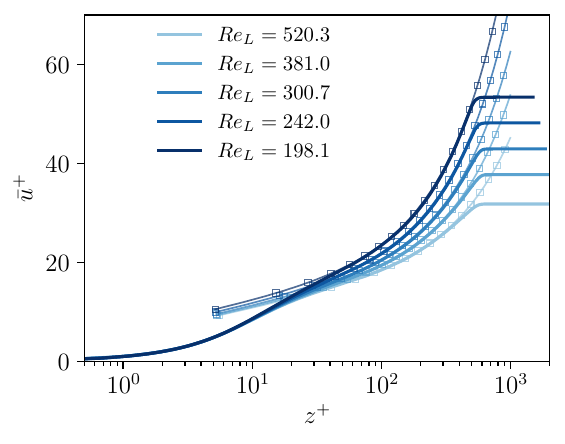}%
    \begin{picture}(0,0)
        \put(-180,140){(a)}
    \end{picture}
\end{minipage}
\hfill
\begin{minipage}[b]{0.48\textwidth}
    \centering
    \includegraphics[width=\linewidth]{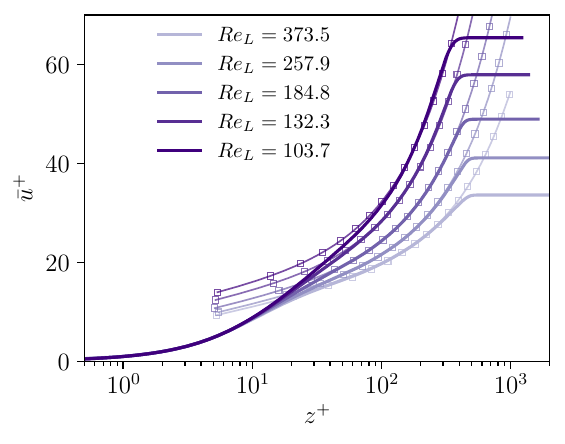}%
    \begin{picture}(0,0)
        \put(-180,140){(b)}
    \end{picture}
\end{minipage}
\caption{Comparison between mean streamwise velocity profiles from DNS (solid line) and log-linear model accounting for intercept correction (lines with square markers) in cases A1 (a) and A2 (b).}
\label{fig:mean_velocity_profile_model}
\end{figure}

In figure \ref{fig:mean_velocity_profile_model}, the mean velocity profiles in cases A1 and A2 have been reconstructed from Eq. \eqref{eq:loglinear_profile_vel}, with the intercept $A$ modelled from Eq. \eqref{eq:A_fit_ReL}. These two cases have been selected as they display an intermediate $Re_L$-range for which $A$ is gradually shifting from the neutral-case reference at early times to a significantly larger value at later times. With the correction proposed for the intercept, the mean velocity profile can be reconstructed in the outer layer almost until the edge of the boundary layer as $Re_L$ decreases. For illustration, Eq.~\eqref{eq:A_fit_ReL} has also been applied at instants of the simulation that were discarded from the regression since they fall below the threshold $Re_L < 150$. However, it should be kept in mind that the existence of a logarithmic region vanishes below that point due to lack of scale separation. 

In order to obtain a complete description of the mean velocity profile, one should connect the log-linear region to the viscous sublayer following the approach of \cite{vandriest1956} by adding a damping function to the mixing-length model close to the wall, and to the far-field velocity at the boundary-layer edge. As shown by different authors \citep{garcia-villalba2011, atoufi2019,greene2024}, stratification effects first suppress the large turbulent structures in the outer layer. Hence, the wake contribution to the mean flow, introduced in the seminal work of \cite{coles1956}, is rapidly diminished once ground cooling begins to restrict the size of the eddies. Instead, a damping mechanism is introduced near the boundary-layer edge to limit the linear growth due to MOST and improve agreement with the outer-layer velocity. This is accomplished by gradually reducing the slope of the MOST contribution towards zero as one approaches the boundary-layer edge. Close to $\eta = {z}/{\delta} = 1$, the most important contribution to $\bar u^+$ is the linear term arising from MOST. Therefore, inspired by the work of \cite{ding2024}, we define the modified velocity gradient arising from MOST, 
\begin{equation}
\frac{d\Psi_m}{dz} = 0.5\left(1 - \mathrm{erf}\left(\frac{z-z_m}{s}\right)\right)\frac{\beta_m }{\kappa L},
\label{eq:modified_slope}
\end{equation}
where $z_m$ and $s$ control the distance from the wall and the width of the transition region where the slope gradually decreases from $\beta_m/(\kappa L)$ to 0. An important difference between the approach of \cite{ding2024} and the present work is that we observed MOST to be valid almost until the edge of the boundary layer, while these authors assumed MOST to be valid only in the surface layer. Therefore, the purpose of the damping function in the present work is to approximate the velocity near the boundary-layer edge, and $z_m$ scales with $\delta$ (see hereafter), while \citet{ding2024} chose to scale $z_m$ with the Obukhov length by prescribing $z_m = 0.2 L$. 
The blended velocity profile is then reconstructed by integrating Eq. \eqref{eq:modified_slope} over the thickness of the boundary layer and by adding the logarithmic part,
\begin{equation}
    \bar u^+ = \frac{1}{\kappa}\log(z^+) + \frac{\beta_m}{\kappa L} F(z,z_m,s) + A(Re_L), 
    \label{eq:modified_loglin_profile}
\end{equation}
where $F(z, z_m, s)$ is the integral of the damping function, which has a known closed form: 
\begin{align}
    F(z, z_m,s)&  = \int_{z_0}^z 0.5\left(1 - \mathrm{erf}\left(\frac{z'-z_m}{s}\right)\right) \, dz' \nonumber \\
    & = 0.5\left[(z'-z_m) \left(1 - \mathrm{erf}\left(\frac{z'-z_m}{s}\right)\right) - \frac{s}{\sqrt{\pi}} \exp\left(- \left(\frac{z'-z_m}{s} \right)^2\right)\right]_{z_0}^{z}.
    \label{eq:integral_damping}
\end{align}
Since $z_0 \ll z_m$, the lower bound of the integral in Eq. \eqref{eq:integral_damping} can be approximated to zero for the sake of simplicity. Well below the transition region, where $(z_m-z)/s \gg 1$ and the lower integration bound is approximated by zero, $F(z)$ reduces to $z$ and we recover the classical log-linear velocity profile. 

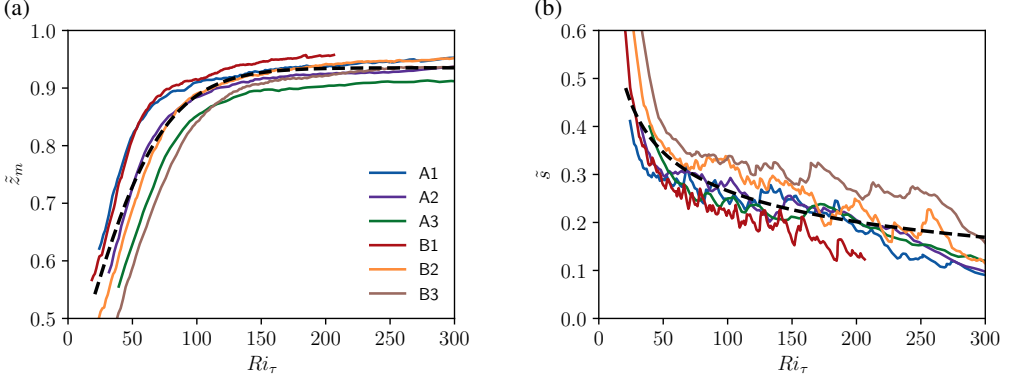
\begin{figure}
\begin{minipage}[b]{0.48\textwidth}
    \centering
    \includegraphics[width=\linewidth]{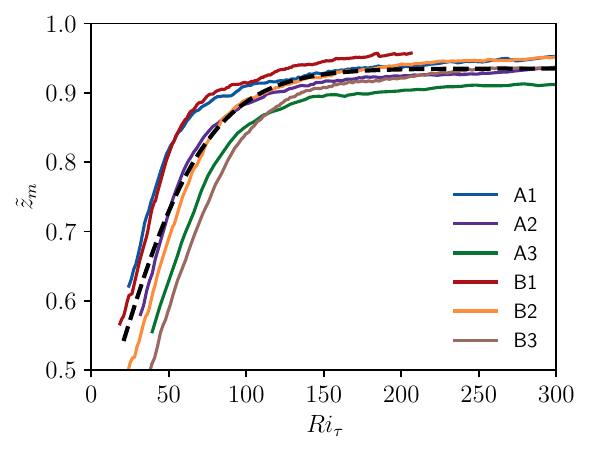}%
    \begin{picture}(0,0)
        \put(-180,140){(a)}
    \end{picture}
\end{minipage}
\hfill
\begin{minipage}[b]{0.48\textwidth}
    \centering
    \includegraphics[width=\linewidth]{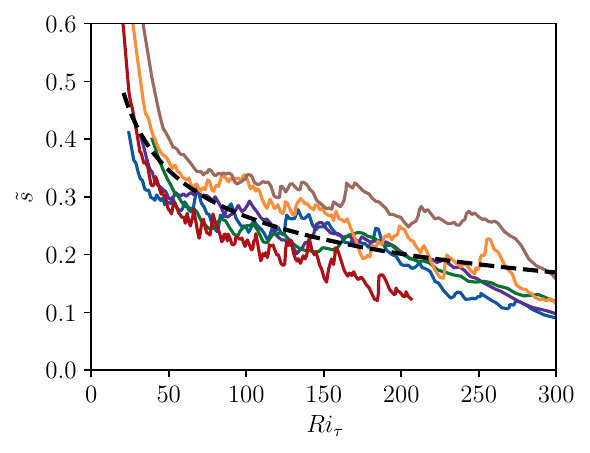}%
    \begin{picture}(0,0)
        \put(-180,140){(b)}
    \end{picture}
\end{minipage}
\caption{Evolution of optimal damping parameters $\tilde z_m$ (a) and $\tilde s$ (b) with $Ri_\tau$ for the six cases A1 to B3. Black dashed lines correspond to the regressions given by Eqs. \eqref{eq:zm_fit} and \eqref{eq:sm_fit}.} 
\label{fig:fit_zm_sm}
\end{figure}

The two parameters $z_m$ and $s$ that control the damping of the slope of $\Psi_m$ can be scaled by the boundary-layer thickness as $\tilde z_m = z_m/\delta$ and $\tilde s = s/\delta$. For each case A1 to B3 and for each time step, optimal $\tilde z_m$ and $\tilde s$ values have been computed to minimize the error (in the least-squares sense) between the DNS profile and the model given by Eq. \eqref{eq:modified_loglin_profile} over the range $z/\delta \in [0.7 , 1.1]$. The upper bound of this range is fixed slightly above 1.0 to include the flattening of the DNS velocity profile beyond $z=\delta$. Since the logarithmic contribution remains undamped, Eq.~\eqref{eq:modified_loglin_profile} is an approximation over the fitted outer-layer range, rather than a composite profile satisfying the viscous-sublayer and far-field limits. Figure \ref{fig:fit_zm_sm} presents the evolution of the fitted $\tilde z_m$ and $\tilde s$ for the six cases as a function of $Ri_\tau$. Despite some scatter between cases, a few general conclusions can be drawn. On the one hand, $\tilde z_m$ increases at early times (i.e. low stratification levels), %
before saturating just below 1, regardless of the case being considered. On the other hand, $\tilde s$ increases towards low $Ri_\tau$ and decreases overall as $Ri_\tau$ increases. To model the log-linear and outer portions of the velocity profile, we consider the following expressions for $\tilde z_m$ and $\tilde s$ as a function of $Ri_\tau$. The black dashed lines in figure \ref{fig:fit_zm_sm} are given by the following expressions: 
\begin{equation}
    \tilde z_m = \tilde z_{m,\infty} - a_z \exp(-c_z  Ri_\tau^{p_z}),
    \label{eq:zm_fit}
\end{equation}
where $\tilde z_{m,\infty} = 0.935$, $a_z = 0.521$, $p_z = 1.362$ and $c_z = 4.49  \times 10^{-3}$, 
and 
\begin{equation}
    \tilde s = \tilde s_{\infty} + \frac{a_s}{Ri_\tau^{p_s}} \exp(-c_s \, Ri_\tau),
    \label{eq:sm_fit}
\end{equation}
where $\tilde s_{\infty} = 0.0528$, $a_s = 1.473$, $p_s = 0.402$ and $c_s = 8.23 \times 10^{-4}$. The power-law term in the denominator of Eq. \eqref{eq:sm_fit} ensures that $\tilde s$ tends to infinity as $Ri_\tau$ vanishes. The damping factor in Eq.~\eqref{eq:modified_slope} then tends to $1/2$ at fixed $z/\delta$, while the stratification contribution  vanishes as $L \rightarrow \infty$. We should note that, to obtain a general expression for the mean velocity profile valid from the neutral to the very stable regime, Coles' law of the wake \citep{coles1956} should be incorporated into  Eq.~\eqref{eq:modified_loglin_profile}, in a form that ensures that its contribution is negligible as $Ri_\tau$ increases. This further generalization is left for future work.

\begin{figure}
\begin{minipage}[b]{0.48\textwidth}
    \centering
    \includegraphics[width=\linewidth]{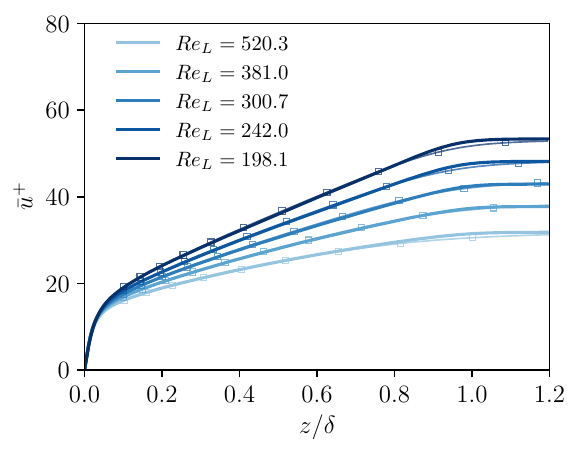}%
    \begin{picture}(0,0)
        \put(-180,140){(a)}
    \end{picture}
\end{minipage}
\hfill
\begin{minipage}[b]{0.48\textwidth}
    \centering
    \includegraphics[width=\linewidth]{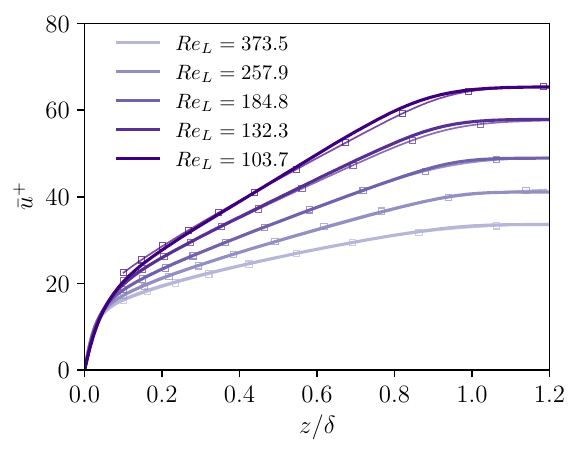}%
    \begin{picture}(0,0)
        \put(-180,140){(b)}
    \end{picture}
\end{minipage}
\caption{Comparison between DNS mean velocity profiles (solid line) and the damped log-linear model given by Eq. \eqref{eq:modified_loglin_profile} (lines with square markers) in cases A1 (a) and A2 (b).}
\label{fig:mean_velocity_profile_model_damping}
\end{figure}
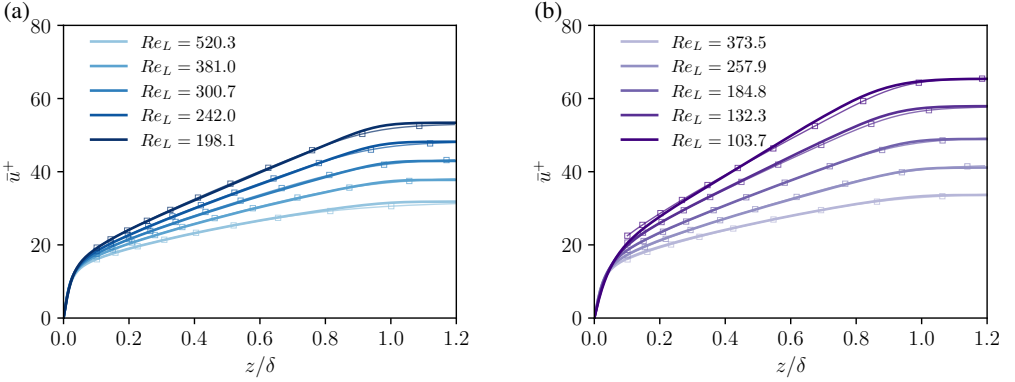

Figure \ref{fig:mean_velocity_profile_model_damping} compares DNS velocity profiles in cases A1 and A2 against the model proposed in Eq. \eqref{eq:modified_loglin_profile}, with the parameters $\tilde z_m$ and $\tilde s$ computed from Eqs.~\eqref{eq:zm_fit} and \eqref{eq:sm_fit}. The comparison shows good agreement over the log-linear and outer-layer regions and for a wide range of stratification levels.  

\subsection{Wall friction coefficient }
\begin{figure}
\begin{minipage}[b]{0.48\textwidth}
    \centering
    \includegraphics[width=\linewidth]{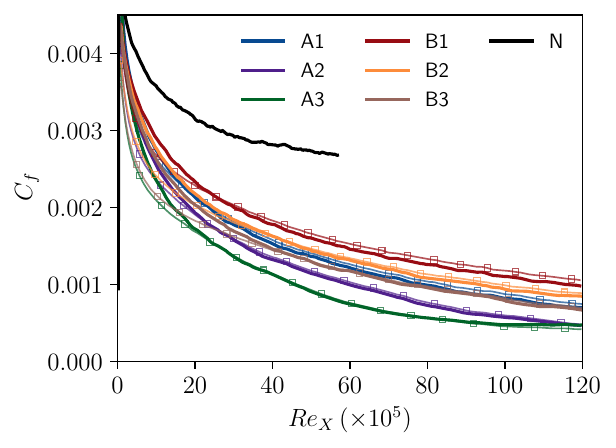}%
    \begin{picture}(0,0)
        \put(-180,140){(a)}
    \end{picture}
\end{minipage}
\hfill
\begin{minipage}[b]{0.48\textwidth}
    \centering
    \includegraphics[width=\linewidth]{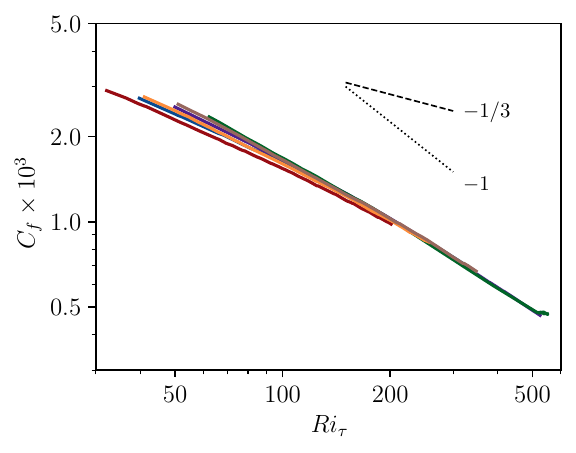}%
    \begin{picture}(0,0)
        \put(-180,140){(b)}
    \end{picture}
\end{minipage}
\caption{Evolution of the wall friction coefficient $C_f$ with $Re_X$ (a) and $Ri_\tau$ in log-log scale (b). In (a), DNS data for cases A1 to B3 (solid lines) are compared to the model given by Eq. \eqref{eq:Uw_plus_model} (lines with square markers). In (b), the -1/3 and -1 power laws are indicated by dashed and dotted lines, respectively.}
\label{fig:Cf_time_evol}
\end{figure}

Now that we are equipped with a model for the mean streamwise velocity profile accounting for stratification effects in the log-linear and outer regions, the wall friction coefficient $C_f$ can be obtained from the evaluation of the mean velocity profile at $z=\delta$ as $C_f = 2\tau_w/(\rho U_w^2) = 2u_\tau^2/U_w^2 = 2/(U_w^+)^2$, where $U_w^+$ is approximated by $\bar u_\delta^+ = \bar u^+(z=\delta)$. %
With this approximation,
\begin{equation}
    U_w^+ \simeq \bar u_\delta^+= \frac{1}{\kappa}\log \left(Re_\tau \right) + \frac{\beta_m}{\kappa L} F_\delta + A(Re_L).
    \label{eq:Uw_plus_model}
 \end{equation}
 where $F_\delta = F(\delta, z_m(Ri_\tau, \delta), s(Ri_\tau, \delta))$. 
The evaluation of $C_f$ from DNS and from the model given by Eq.~\eqref{eq:Uw_plus_model} are compared in figure \ref{fig:Cf_time_evol}(a). After a transient period ($Re_X \lesssim 30 \times 10^5$), the agreement between the data and the model is extremely good. The temporal evolution of $C_f$ in the neutral case is also included, confirming the significant reduction in wall shear stress under stable stratification, consistent with the conclusions of previous experimental \citep{arya1975} and numerical \citep{garcia-villalba2011,zonta2013} studies. 

Figure \ref{fig:Cf_time_evol}(b) shows the evolution of the friction coefficient with $Ri_\tau$ for the six stratified cases in log-log scale. %
\cite{garcia-villalba2011} reported a scaling $C_f \propto Ri_\tau^{-1/3}$ from their DNS database of stratified channel flows. This trend was confirmed by \cite{zonta2018} in their review study on stratified wall-bounded turbulence. However, most studies selected in this review concerned stably stratified channel flows. In the present case, we observe a gradual change in the apparent power law from $Ri_\tau^{-1/3}$ to $Ri_\tau^{-1}$ as stability increases. The definitions give the exact identity $C_f Ri_\tau = 2 Ri_\delta$. Since $Ri_\delta$ is proportional to $\delta$ within each run, the approach towards a $-1$ slope is consistent with boundary-layer growth slowing relative to the increase in $Ri_\tau$.

\section{Conclusion}

\label{section:conclusion}
The flow in a stably stratified temporally developing turbulent boundary layer (TTBL) has been investigated using direct numerical simulation over a range of Richardson and Reynolds numbers. %
We first examined the temporal evolution of key flow quantities. The progressive attenuation and eventual collapse of turbulence observed under strong stratification were quantified and compared against several stability criteria proposed in the literature, including the Reynolds number based on the Obukhov length ($Re_L$) and the bulk Richardson number ($Ri_\delta$). In case A3, we found the onset of turbulence collapse to be consistent with the $Re_L \simeq 100$ threshold of \cite{flores2011} and to coincide with the emergence of intermittent laminar--turbulent patches. 

The DNS mean velocity profiles showed that Monin–Obukhov similarity theory (MOST) remains remarkably well satisfied in the stratified TTBL. The DNS profiles are consistent with the classical similarity coefficients reported in atmospheric observations and laboratory experiments, despite the idealized and non-stationary nature of the present configuration. The local scaling theory introduced by \cite{nieuwstadt1984} was shown to provide the most robust description of the flow across first- and second-order statistics, although the global (surface-based) similarity theory accurately predicts the mean velocity gradient over a large portion of the boundary layer. We attribute this to the similar decay rate of turbulent shear stress and turbulent heat flux with distance from the wall. This conclusion should, however, be restricted to the type of stratified turbulent boundary layer investigated in this study, namely neglecting flow rotation and in the absence of any top-down turbulent process.

Then, we systematically compared the log-linear velocity profile arising from MOST to the DNS data. As stratification increases, the intercept $A$ of the velocity profile shows a clear upward shift, which we attribute to a reduction of the inner-scaled shear stress in the near-wall region. The deviation of $A$ with respect to its neutral-case value correlates well with the Reynolds number based on the Obukhov length, providing a physically meaningful parametrization of stratification effects on the mean flow near the wall.

Finally, a correction to the mean velocity profile was proposed to improve agreement with the DNS velocity near the boundary-layer edge. This was achieved by applying a damping function to the linear stratification term arising from MOST. The resulting velocity model yields accurate predictions of the wall friction coefficient, which compares favourably with the DNS data across all cases considered. %

Overall, the present study shows that the temporally developing boundary layer is a simple, tractable, yet physically faithful simulation framework to systematically study the applicability of scaling theories in turbulent boundary layers with complex physics, such as thermal stratification, and to build new closures that can serve as a basis for improved wall models. %

\backsection[Acknowledgements]{We acknowledge the use of the Dutch national supercomputer Snellius under NWO Grant no.\ \texttt{2024/ENW/01704792}. We also acknowledge EuroHPC for access to the supercomputer Leonardo, hosted at CINECA (Italy), through Grant no.\ \texttt{EHPC-EXT-2022E01-054}. Finally, we thank Sanath Kotturshettar for the numerous fruitful discussions about Monin--Obukhov similarity theory and its implications.}

\backsection[Funding]{This work was funded by a  Wallonia-Brussels International (WBI) Postdoctoral Excellence Scholarship for the academic years 2023--2024 and 2024--2025.}

\backsection[Declaration of interests]{The authors report no conflict of interest.}

\backsection[Data availability statement]{Data supporting the findings of this study can be shared upon request.}

\appendix
\section{Richardson number definitions in stratified boundary layers}
\label{appendix:richardson}
The relative importance of stratification effects within a turbulent boundary layer is best quantified by the flux Richardson number, a local ratio between buoyancy-induced destruction of turbulent kinetic energy (in the stable regime) and production by shear: 
\begin{equation}
    Ri_f = \frac{\beta g \, \overline{w'\Theta'}}{\overline{u'w'}\dfrac{\partial \bar u}{\partial z}}.
    \label{eq:Rif}
\end{equation}
In practice, the Richardson number based on the mean velocity and temperature gradients, $Ri_g$, is more easily measured from field data and can be related to the flux Richardson number as 
\begin{equation}
    Ri_g = \frac{\beta g \dfrac{\partial \bar \Theta}{\partial z}}{\left(\dfrac{\partial \bar u}{\partial z}\right)^2} = Pr_t Ri_f, 
    \label{eq:Rig}
\end{equation}
where $Pr_t = \dfrac{K_M}{K_H}$ is the turbulent Prandtl number, and $K_M$ and $K_H$ are the turbulent eddy viscosity and turbulent thermal diffusivity, respectively. 
From Eqs. \eqref{eq:ObukhovL}, \eqref{eq:MOST} and \eqref{eq:Rif}, it follows that the flux Richardson number and $\zeta$ are connected in the constant-flux surface layer  by $Ri_f = \zeta/ \phi_m(\zeta) $. Moreover, it can be shown that $Pr_t = \phi_h/\phi_m$, which finally yields
\begin{equation}
    Ri_g = \zeta \frac{\phi_h(\zeta)}{\phi_m^2(\zeta)}.
    \label{eq:Rig_zeta}
\end{equation}
Hence, the original formulation of MOST based on $\zeta$ is also unambiguously related to Richardson-based formulations proposed in the literature \citep{grachev2013, sorbjan2010}.

\section{Scaling of the mean temperature profile}
\label{appendix:temperature}

\begin{figure}
\begin{minipage}[b]{0.48\textwidth}
    \centering
    \includegraphics[scale=.62]{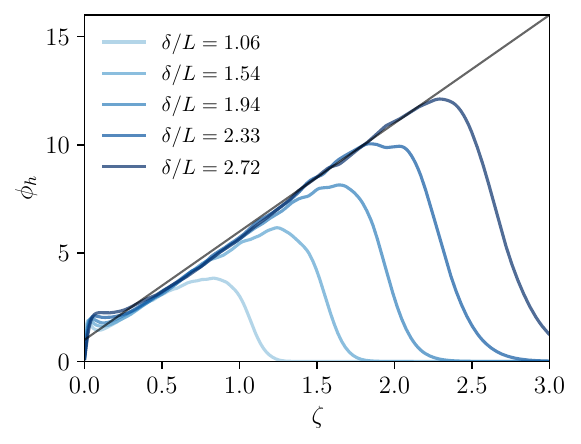}
    \begin{picture}(0,0)
        \put(-180,130){(a)}
    \end{picture}
    \label{fig:mean_temp_profile_a}
\end{minipage}
\hfill
\begin{minipage}[b]{0.48\textwidth}
    \centering
    \includegraphics[scale=.62]{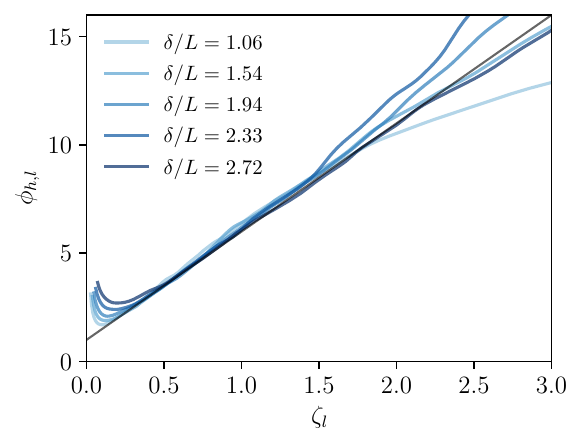}
    \put(-180,130){(b)}
    \label{fig:mean_temp_profile_b}
\end{minipage}

\begin{minipage}[b]{0.48\textwidth}
    \centering
    \includegraphics[scale=.62]{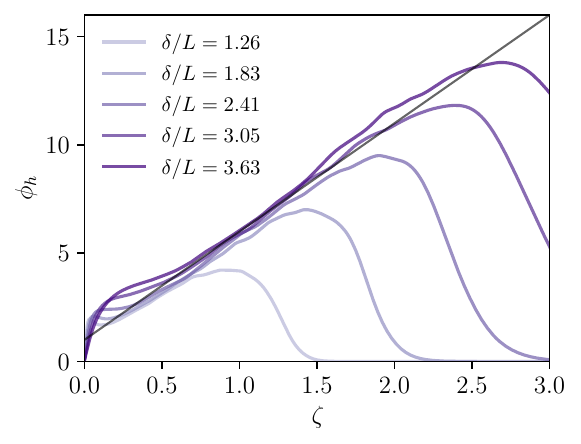}
    \put(-180,130){(c)}
    \label{fig:mean_temp_profile_c}
\end{minipage}
\hfill
\begin{minipage}[b]{0.48\textwidth}
    \centering
    \includegraphics[scale=.62]{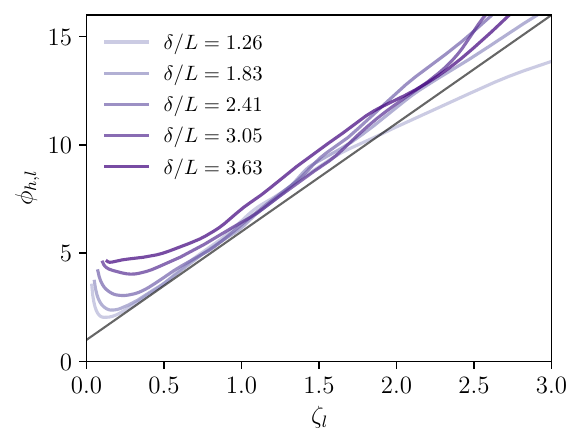}
    \put(-180,130){(d)}
    \label{fig:mean_temp_profile_d}
\end{minipage}

\begin{minipage}[b]{0.48\textwidth}
    \centering
    \includegraphics[scale=.62]{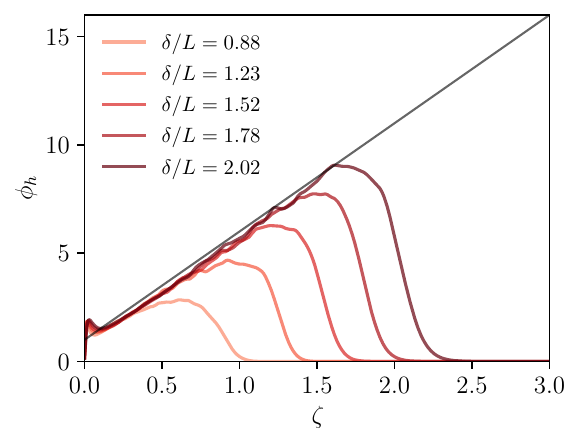}
    \put(-180,130){(e)}
    \label{fig:mean_temp_profile_e}
\end{minipage}
\hfill
\begin{minipage}[b]{0.48\textwidth}
    \centering
    \includegraphics[scale=.62]{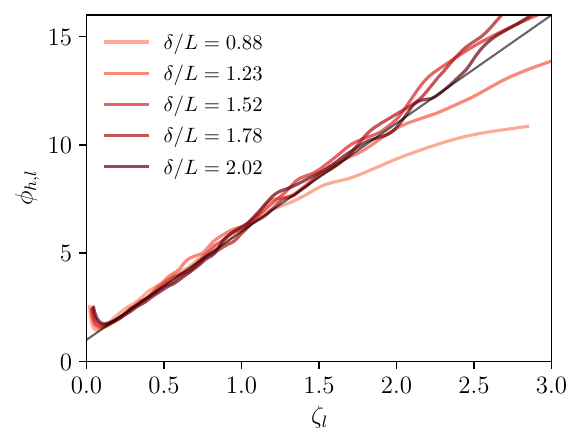}
    \put(-180,130){(f)}
    \label{fig:mean_temp_profile_f}
\end{minipage}
\caption{Mean temperature gradient in global scaling ((a), (c), (e)) and local scaling ((b), (d), (f)) at different instants of the flow in case A1 ((a), (b)), A2 ((c), (d)) and B1 ((e), (f)). DNS results are compared with the similarity relation $\phi_h = 1 + 5.0 \zeta$ (black solid line) \citep{dyer1974, brutsaert1982}. In the local panels, the linear relation is $\phi_{h,l}=1+5.0\zeta_l$.}
\label{fig:phi_h}
\end{figure}
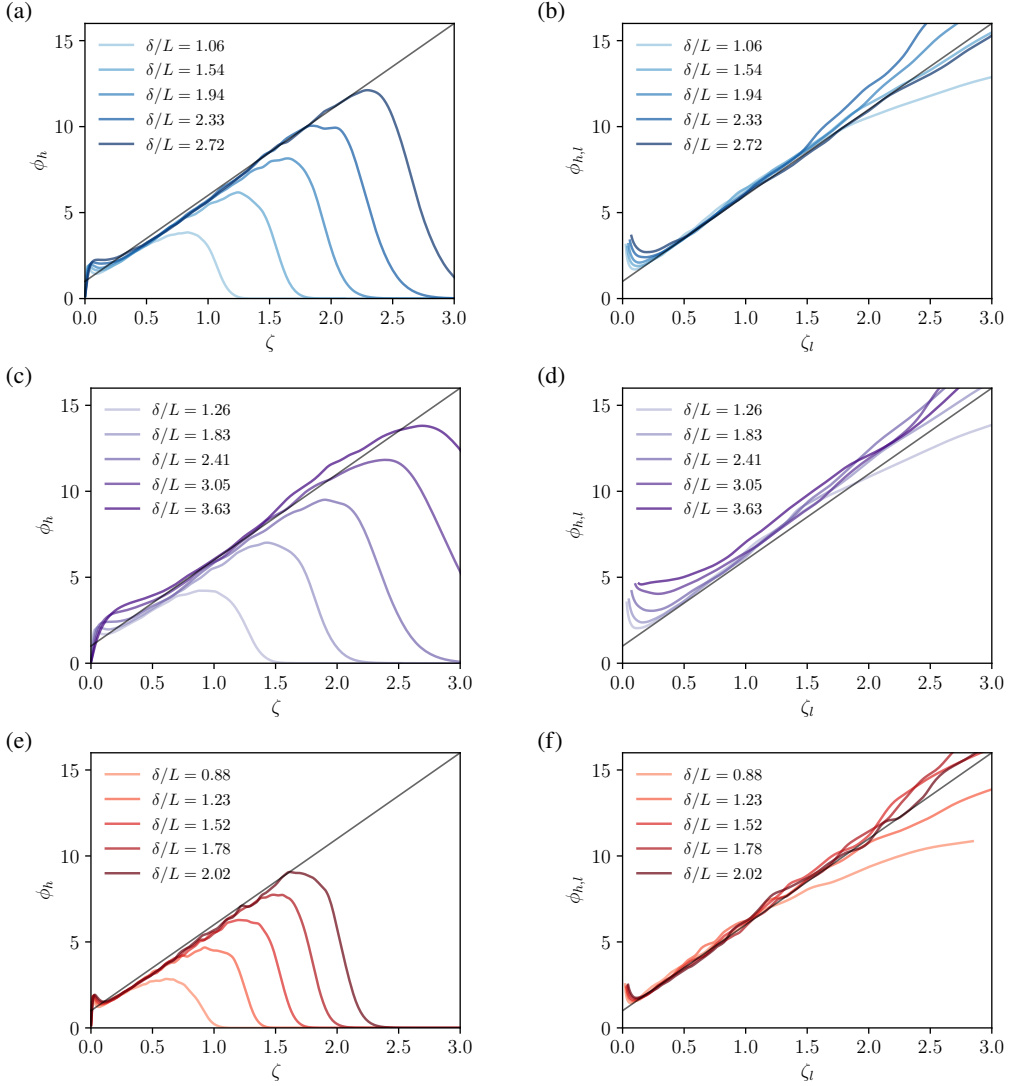

For completeness, we also assess from DNS data the accuracy of MOST for the temperature gradient profile in the stably stratified TBL. Figure \ref{fig:phi_h} shows profiles of the temperature gradient $\phi_h$ in global and local scaling, for cases A1, A2 and B1 and different times. Key conclusions drawn from the velocity gradient profiles in Section \ref{section:similarity_theory} extend to the temperature field: departure from the linear similarity relation is controlled by proximity to the boundary-layer edge rather than by 
$\zeta$ itself, and the extended range of validity of global scaling is consistent with the similar decay rates of turbulent shear stress and heat flux discussed above.

The DNS profiles are compared in figure \ref{fig:phi_h} with the linear relation given by Eq. \eqref{eq:phi_h} and coefficients $\alpha_h=Pr_{t_0} =1$ and $\beta_h=5.0$, which amounts to setting $\phi_h = \phi_m$, as suggested by \cite{dyer1974}. The agreement between the DNS data and the similarity relation appears very good, both in global and local scaling. 
The intercept $\alpha_h$ estimated directly from the linear-regression fit of Eq. \eqref{eq:phi_h} is, however, highly sensitive to the interval of wall-normal positions retained for the regression. A more robust estimate can instead be obtained from the slope of the log-linear region of the inner-scaled mean temperature profile in the stratified cases, $1/\kappa_\theta$, from which the turbulent Prandtl number follows as $Pr_t = \frac{\kappa}{\kappa_\theta}$. This diagnostic yields $Pr_t \simeq 0.85$ within the log-region, though this value remains sensitive to the stratification level across our stratified cases. 

\clearpage
\bibliographystyle{jfm}
\bibliography{biblio}

\end{document}